\documentclass{nature}
\usepackage{amsmath}
\usepackage{amsfonts}
\usepackage{bm}
\usepackage{xcolor}
\usepackage{graphicx}
\usepackage{caption}

\newcommand{\ui}{\mathrm{i}}
\newcommand{\ud}{\mathrm{d}}
\newcommand{\be}[0]{\begin{equation}}
\newcommand{\ee}[0]{\end{equation}}
\newcommand{\bea}[0]{\setlength\arraycolsep{2pt}\begin{eqnarray}}
\newcommand{\eea}[0]{\end{eqnarray}}
\newcommand{\bse}{\begin{subequations}}
\newcommand{\ese}{\end{subequations}}
\newcommand{\brho}{\boldsymbol{\rho}}
\newcommand{\bu}{{\bf u}}

\title{Birefringent Fourier filtering for single molecule Coordinate and Height super-resolution Imaging with Dithering and Orientation}

\author{Valentina Curcio$^{1,**}$, Luis A. Alem\'an-Casta\~neda$^{1,2,**}$, Thomas G. Brown$^2$, Sophie Brasselet$^{1,*}$ \& Miguel A. Alonso$^{1,2,*}$}

\usepackage{lineno}

\begin{document}

\maketitle

\begin{affiliations}
 \item Aix Marseille Univ, CNRS, Centrale Marseille, Institut Fresnel, F-13013 Marseille, France
 \item The Institute of Optics, University of Rochester
Rochester, NY 14627, U.S.A.
 \item[**] These authors contributed equally to the work
	\item[*] Corresponding authors: sophie.brasselet@fresnel.fr, miguel.alonso@fresnel.fr
\end{affiliations}

\begin{abstract}
Super-resolution imaging based on single molecule localization allows accessing nanometric-scale information in biological samples with high precision. However, complete measurements including molecule orientation are still challenging. Orientation is intrinsically coupled to position in microscopy imaging, and molecular wobbling during the image integration time can bias orientation measurements. Providing 3D molecular orientation and orientational fluctuations would offer new ways to assess the degree of alignment of protein structures, which cannot be monitored by pure localization. Here we demonstrate that by adding polarization control to phase control in the Fourier plane of the imaging path, all parameters can be determined unambiguously from single molecules: 3D spatial position, 3D orientation and wobbling or dithering angle. The method, applied to fluorescent labels attached to single actin filaments, provides precisions within tens of nanometers in position and few degrees in orientation.
\end{abstract}

\section*{Introduction}
Biological functions in cells and tissues are driven by the molecular-scale organization of biomolecular assemblies, which arrange in precise structures that are essential, for instance, in biomechanics and morphogenesis. A way to assess such organization is to monitor the orientation of fluorescent labels, in conditions where the label is sufficiently rigidly attached to the biomolecule of interest\cite{Beausang2013,ValadesCruz2016,Mehta2017,Ding2020}.  Monitoring orientational behavior of fluorescent molecules is still a challenge, however, because both orientational fluctuations and mean orientation need to be quantified. In particular, measurements can be strongly biased by the fact that molecular orientations may fluctuate at a time scale faster than the measurement integration time, which occurs naturally in biological media even in fixed conditions\cite{Backer2015,ValadesCruz2016,Ding2020}. Recent studies have aimed at adding  orientation information to super-resolution imaging, which relies on single molecule localization. Orientation and position are however difficult parameters to disentangle, leading to possible localization biases\cite{Enderlein2006,Backlund2012}. A single molecule's point spread function (PSF) is intrinsically altered by its orientational properties\cite{Enderlein2006,Ding2020}. Several methods have capitalized on this property by using Fourier-plane phase modification of the PSF\cite{Backlund2012,Agrawal,Backer2013,Backer2014}, or imaging finely sampled PSFs\cite{Mortensen2010}. However, these approaches apply only to molecules with fixed orientation. 
Recent proposals to access the missing information on wobbling rely on adding complexity to the PSF via phase filtering\cite{Zhang2018} or by using the index mismatch sensitivity of the PSF's shape\cite{Ding2020}, although the axial component of the single molecules' 3D position remains inaccessible. 
Other approaches use defocused imaging\cite{Aguet2009,Backer2015}, but they require either fixed orientations or predetermined spatial localization of the molecules. Alternatively, it is possible to preserve less-altered PSF images and restrict the measurements to 2D in-plane orientations by working under relatively low numerical aperture conditions and splitting polarization components\cite{ValadesCruz2016}, or using sequential polarization illumination\cite{Corrie1999,Sosa2001,Peterman2001,Backer2016}. So far, none of these techniques have allowed the simultaneous measurement of 3D orientational properties (including both orientational fluctuations and mean orientation) and 3D spatial position of single molecules, in a single-shot image scheme compatible with super-resolution localization. The main challenge is that the axial position of single molecules and their 3D orientational fluctuations (e.g. their wobbling) are intrinsically coupled by the imaging techniques.

\begin{figure}
\centering
\includegraphics[width=0.9\linewidth]{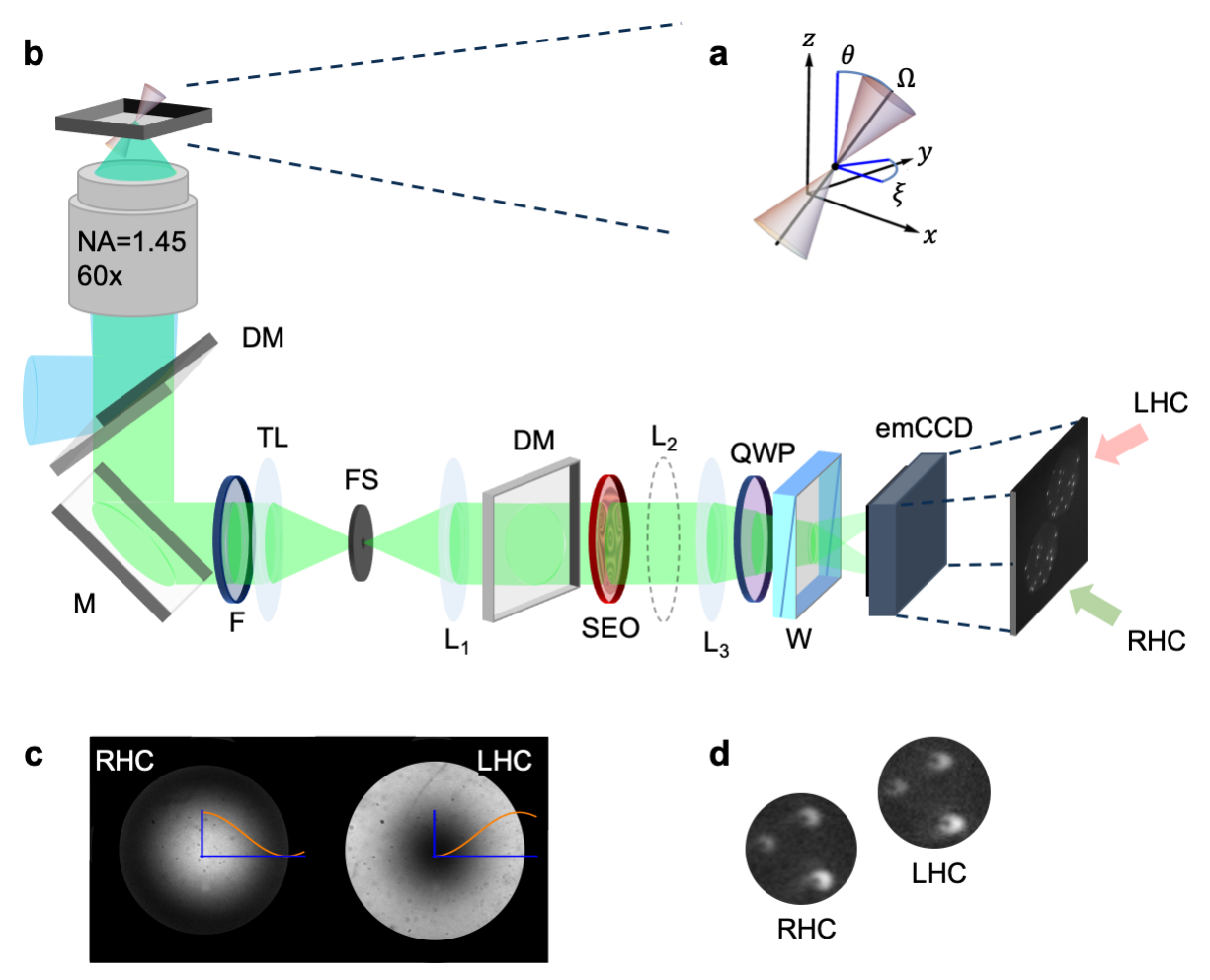}
\caption{CHIDO imaging principle. (a) Parameters defining the 3D position $(x,y,z)$, orientation $(\xi,\theta)$ and wobbling subtended solid angle $\Omega$.  (b) Optical setup (see Methods).  
DM: dichroic mirror. M: mirror. F: fluorescence filter. TL: tube lens. FS: field stop. L$_1$, L$_2$ and L$_3$: lenses. QWP: quarter wave plate. W: Quartz Wollaston polarizing beamsplitter. emCCD: emCCD camera. RHC and LHC: right-hand circular and left-hand circular polarized images. (c) Back focal plane imaging of the SEO illuminated by a circular polarization (the sample is a homogeneous fluorescent sample; see Methods). (d) Direct image of isolated emitters (fluorescence beads, see Section 4) in the same polarized emission conditions.
}
\label{setup}
\end{figure}
Here we propose a simple method to engineer the molecular PSF so that it efficiently encodes information about all these properties with very little coupling. The method is based on Fourier-plane filtering not only in phase but also in polarization by using spatially-varying birefringence. It builds upon a prior technique for single-shot imaging polarimetry\cite{Roshita,Brandon,Brandon2}, where polarization is encoded in the shape of the PSF. This approach has been applied to 
multiple scattering measurements\cite{BrandonAerosols} as well as to the polarimetric characterization of multicore fibers\cite{Sid}. In this work, we show that the same operating principle can be used to retrieve significantly more degrees of freedom when applied to imaging fluorescing molecules, where the PSFs encode information not only of the molecules' transverse coordinates $(x,y)$ but also of their axial height $z$, and of the three-dimensional correlations of the emitted light, which translate into the orientation of the molecules, namely the azimuthal angle $\xi$, the polar angle $\theta$, and the state of wobbling or dithering  characterized by the average cone solid angle $\Omega$ (Fig.~\ref{setup}(a)). 
Furthermore, we show that there is negligible coupling in the dependence of the PSFs on the relevant parameters being measured, that the technique involves almost no photon losses, and that transverse spatial resolution is high since the PSFs encoding this information are only about twice as large as those of diffraction-limited imaging. We refer to the method as Coordinate and Height super-resolution Imaging with Dithering and Orientation (CHIDO).

\section*{Results}
\subsection{PSF encoding through a birefringent mask.}
The basis of the proposed technique is the placement at the pupil plane of an element referred to as a stressed-engineered optic (SEO), which is a BK7 glass window subjected to forces with trigonal symmetry at its edges\cite{Alexis,Roshita,Brandon} (see Methods). The spatially-varying birefringence pattern that naturally results in the vicinity of the force equilibrium point has been shown to be essentially optimal for applications in polarimetry, in the sense that it efficiently encodes polarization information in the PSF's shape while causing the smallest possible increase in PSF size\cite{Anthony}. This birefringence pattern is described by the following Jones matrix (in the linear polarization basis) in the Fourier plane of the detection path (Fig.~\ref{setup}(b)):
\be\label{JonesMatrixSEO}
\mathbb{J}(\bu)\!=\!\cos\!\frac{cu}2\left(\!\begin{array}{cc}\!1&0\\0&1\end{array}\!\right)+\ui\,\sin\!\frac{cu}2\left(\!\begin{array}{cc}\cos\varphi&\!\!-\sin\varphi\\-\sin\varphi&\!\!-\cos\varphi\end{array}\!\right),
\ee
where $(u,\varphi)$ are polar pupil coordinates normalized so that $u=1$ corresponds to the pupil's edge, and $c$ is a coefficient that depends on the stress within the SEO and the radius of the pupil being used. This parameter can be chosen to optimize the system's performance: small $c$ keeps the extension of the PSFs more restricted but reduces the amount of information they carry about orientation and $z$ displacement, while large $c$ has the opposite effect\cite{Roshita,Anthony}.
After passing through the SEO, the two circular polarization components are separated to form two images by inserting a quarter-wave plate (QWP) followed by a Wollaston prism (Fig.~\ref{setup}(b)). A Fourier-plane image under circularly-polarized illumination shows the effect of the SEO's spatially-varying birefringence as a Fourier mask on the two detection channels (Fig.~\ref{setup}(c)).

We now show that the combination of the SEO and the separation of the two circular polarization images allows encoding information about a molecule's orientation and axial displacement in the shape of the PSFs. Let us model the fluorescing molecule as a quasi-monochromatic point dipole that can have any orientation (fixed or fluctuating) in three dimensions\cite{Bohmer:03,Aguet2009,Backer2015,HieuThao:20,Chandler1:19,Chandler2:19}. For now we assume that this dipole is at the center of the object focal plane of the objective, $(x,y,z)=(0,0,0)$; the effects of lateral and axial displacements will be discussed later. The dipole is placed in a homogeneous medium, at a distance to the glass coverslip larger than the wavelength. This source can be described by the $3\times3$ second moment (or correlation) matrix $\mathbf{\Gamma}$ with elements $\Gamma_{ij} =\langle E_i^*E_j\rangle$ with $i,j=x,y,z$, $E_i$ being the radiated field components, and the angular brackets denoting an average over the integration time of the detector\cite{Backer2015} (Supplementary Note 1).
This type of $3\times3$ correlation matrix has also been used to study nonparaxial polarization\cite{Brosseau,Sampson,Barakat,Tero,2methods,Alonso:2020geometric}.  For the sake of analogy with standard polarimetry (where the correlation matrix is only $2\times2$), we write $\mathbf{\Gamma}$ in terms of the  generalized 3D Stokes parameters $S_n$, which are the coefficients of the expansion of this matrix in terms of the Gell-Mann matrices $\mathbf{g}_n$ (instead of the Pauli matrices used for $2\times2$ correlations, whose coefficients are the standard Stokes parameters)\cite{Brosseau}. The resulting expression is
\be
\!\mathbf{\Gamma}\!=\!\sum_{n=0}^8S_n\mathbf{g}_n=\left(\begin{array}{ccc}\frac{S_0+S_8}{\sqrt{3}}+S_1&S_2-\ui S_3&S_4-\ui S_5\\S_2+\ui S_3&\frac{S_0+S_8}{\sqrt{3}}-S_1&S_6-\ui S_7\\ S_4+\ui S_5&S_6+\ui S_7&\frac{S_0-2S_8}{\sqrt{3}}\end{array}\right).
\label{Gamma}
\ee
Note that we use a nonstandard numbering scheme for the Gell-Mann matrices: the elements $n=1,2,3$ are cycled so that the resulting parameters $S_n$ reduce to the standard Stokes parameters for $n=1,2,3$ when the field's $z$ component vanishes. (In this case, $S_0$ differs from the corresponding Stokes parameter for paraxial light by a factor of $\sqrt{3}/2$.) 

Several measures have been proposed for the degree of polarization of nonparaxial light\cite{2methods,Alonso:2020geometric}, one of them\cite{Sampson,Barakat,Tero} having a definition in terms of the generalized 3D Stokes parameters that resembles the standard one for paraxial light:
\be
P_{3{\rm D}}=\frac1{S_0}\left(\sum_{n=1}^8S_n^2\right)^{1/2}=\left[\frac{3\,{\rm tr}\mathbf{\Gamma}^2}{2\,({\rm tr}\mathbf{\Gamma})^2} -\frac12\right]^{1/2}.
\ee
In the present context, this degree of polarization is related to the amount of wobbling of the fluorescent dipole source. For a dipole wobbling uniformly within a cone, the cone solid angle $\Omega$ is a monotonic function of this degree of polarization (see Supplementary Note 2):
\be
\Omega=\pi\left(3-\sqrt{1+8P_{3{\rm D}}}\right)\,\,\,\,\,\,\,{\rm or}\,\,\,\,\,P_{\rm 3D}=\frac{(3\pi-\Omega)^2-\pi^2}{8\pi^2}.\label{Omega}
\ee
Note that, for isotropic wobbling (that is, when the two smallest eigenvalues of $\mathbf{\Gamma}$ are equal), $P_{\rm 3D}$ coincides with the rotational mobility parameter proposed to characterize wobbling\cite{Zhang2019}. The relation between these two measures is described in Supplementary Note 2; these and other related measures of polarization have simple geometric interpretations\cite{Alonso:2020geometric}.

\begin{figure}
\centering
\includegraphics[width=0.9\linewidth]{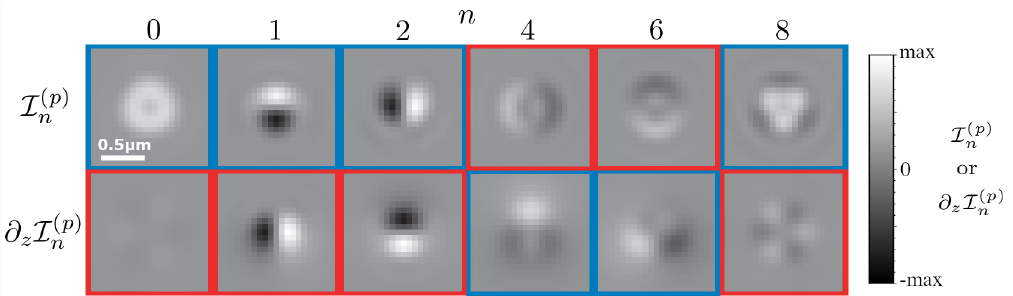}
\caption{Theoretical PSF components in CHIDO imaging. The figure shows both ${\cal I}_n^{(p)}$ and $\partial_z{\cal I}_n^{(p)}$ for $c=\pi$, $z=0$, and $p=\rm \mathcal{R}$. The corresponding components for $p=\rm \mathcal{L}$ are identical, except that those surrounded by red boxes would have the opposite sign. Each row is normalized separately as their units are different.}
\label{PSFcomps}
\end{figure}
Let the PSFs at the two detector regions be denoted as $I^{(p)}$, where $p$ labels the  polarization component being imaged at the corresponding detector: $p=\mathcal{R}$  for right-hand circular (RHC) and $p=\mathcal{L}$  for left-hand circular (LHC) (Fig.~\ref{setup}(d)). As shown in Supplementary Note 1, these PSFs depend linearly on the generalized Stokes parameters according to
\be
I^{(p)}(\brho)=\sum_{n=0}^8S_n{\cal I}_n^{(p)}(\brho),
\label{linearrel}
\ee
where ${\cal I}_n^{(p)}$ are contributions to the PSF corresponding to each generalized Stokes parameter. Expressions for these contributions are derived in Supplementary Note 1, and theoretical images for some of them at $z=0$ are shown in the top row of Fig.~\ref{PSFcomps}. 
Note that Fig.~\ref{PSFcomps} does not include images for  ${\cal I}_3^{(p)}$, ${\cal I}_5^{(p)}$, and ${\cal I}_7^{(p)}$ because they are not of interest to the current problem. (The complete set is shown in Supplementary Figure 1.) This is because, as we can see from Eq.~(\ref{Gamma}), the generalized Stokes parameters $S_3$, $S_5$ and $S_7$ correspond to the imaginary part of $\mathbf{\Gamma}$ and therefore encode information about the helicity of the emitted field, which is assumed not to exist since the emitters are (possibly wobbling) linear dipoles. Nevertheless, if the particles did emit light with some helicity, 
these elements could be incorporated into the treatment.

An important feature of the SEO's birefringence pattern is that it makes this set of PSF components nearly orthogonal while keeping their extension almost as small as possible (in analogy to the case of paraxial polarization\cite{Anthony}). This approximate orthogonality implies the strong decoupling of the information for each parameter, as will be discussed in the next section. Another desirable aspect of using the SEO as a filter at the pupil plane is the resulting approximate achromaticity over a spectral range corresponding to fluorescence spectral widths (typically 100nm), in contrast to PSF engineering methods based on pure phase masks. The only chromatic dependence of the Jones matrix in Eq.~(\ref{JonesMatrixSEO}) is within the parameter $c$, which is roughly inversely proportional to the wavelength. This variation compensates the natural scaling of the PSF with wavelength, such that the PSFs resulting from the integration over the fluorescence spectrum being measured are nearly indistinguishable from those resulting from only the peak wavelength. 
If a measurement required larger wavelength ranges, appropriate recalibration must be used, as in any PSF-engineering-based technique. The chromatic dependence of CHIDO is discussed in Supplementary Note 1 and quantified in Supplementary Figure 2. 

\begin{figure}
\centering
\includegraphics[width=0.5\linewidth]{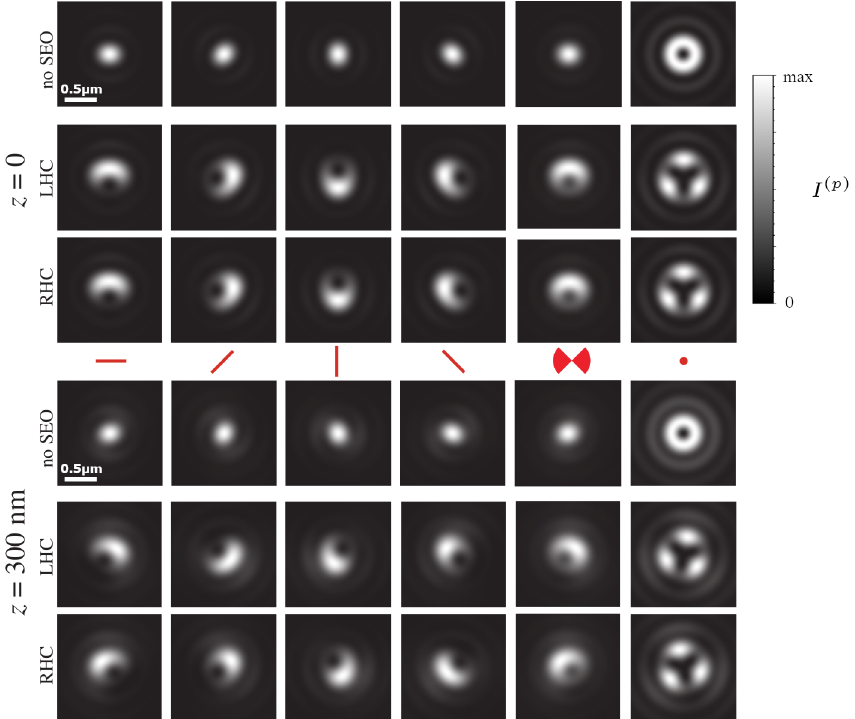}
\caption{Theoretical PSFs formed from specific dipoles orientation and wobbling. The PSFs are shown at the nominal focal plane $z=0$ (top) and at $z=300$ nm (bottom), corresponding to five different dipole orientations: the first four from left to right correspond to non-wobbling dipoles within the $xy$-plane ($\theta=90^\circ$) and for $\xi=0^\circ$, $45^\circ$, $90^\circ$, and $135^\circ$, respectively, while the sixth column corresponds to a non-wobbling dipole in the $z$ direction ($\theta=0^\circ$). The fifth column corresponds again to $\xi=0^\circ$, but for the dipole wobbling within an angle $\delta$ of $90^\circ$ (corresponding to $\Omega=0.6\pi$ and $P_{\rm 3D}=0.6$). 
For both heights, the top row shows for reference the (diffraction-limited) PSFs without the SEO, while the rows labeled RHC and LHC show the two PSFs for CHIDO.  
Note from the first four columns that a rotation of the dipole within the $xy$-plane causes an approximate joint rotation of both PSFs in a direction opposite to that of the dipole and by twice the angle. A change in height, on the other hand, causes approximate rotations of both PSFs in opposite directions with respect to each other. Wobbling causes a blurring of the PSFs. PSF pairs for other orientations and wobbling angles are shown in Supplementary Movie 1.}
\label{PSFexamples}
\end{figure}
When the emitter is within the plane conjugate to the image, each of its two images is a linear combination of the six PSFs shown in the top row of Fig.~\ref{PSFcomps}, according to Eq.~(\ref{linearrel}). The possible differences between the two images arise from a global sign change of two members of the PSF basis set, ${\cal I}_4^{(p)}$ and ${\cal I}_6^{(p)}$. Figure~\ref{PSFexamples} and Supplementary Movie 1 show simulations of measured PSF pairs corresponding to several dipole orientations.
Also shown in Fig.~\ref{PSFexamples} for comparison are the corresponding diffraction-limited PSFs resulting from not using the SEO, whose shape is nearly independent of the in-plane angle $\xi$.
In contrast, when the SEO is used, the PSFs acquire a crescent shape for a dipole within the $xy$-plane, and a rotation of the dipole within this plane results in an approximate rotation of both PSFs, in the opposite sense as the dipole and by twice the angle. Note that these PSFs are only about twice as large as the diffraction-limited ones. A dipole in the $z$ direction, on the other hand, corresponds to a PSF with trigonal symmetry (which is also only about twice as large as the corresponding diffraction-limited PSF). Wobbling of the dipole about its nominal direction has the effect of blurring the PSFs in a predictable way. Therefore, the parameters $S_n$ can be estimated by making the superposition in Eq.~(\ref{linearrel}) agree as closely as possible with the measured pair of PSFs (see Supplementary Note 2). From these  parameters the matrix $\mathbf{\Gamma}$ can be constructed using Eq.~(\ref{Gamma}), which is real and symmetric because $S_3=S_5=S_7=0$. The central direction of the dipole source is then estimated as that of the eigenvector of $\mathbf{\Gamma}$ with the largest eigenvalue. The remaining eigenvectors and eigenvalues provide information about the wobbling of the molecule (Supplementary Note 2).
Additionally, in the minimization procedure that leads to the retrieval of the parameters $S_n$, the transverse $x,y$ position of each emitter can be estimated to within a fraction of a pixel (Supplementary Note 2). 
This analysis can be performed simultaneously for multiple emitters within an image, as long as their PSFs do not overlap.

In addition to orientation and transverse localization, the measured images provide information about axial localization, since the PSFs depend on $z$ (significantly more so than those without the SEO). As shown in Fig.~\ref{PSFexamples} and Supplementary Movie 1, a variation in $z$ for a dipole oriented within the $xy$-plane causes a rotation of both measured PSFs, but these rotate in opposite directions. This is in contrast with an in-plane rotation of the dipole (a change in $\xi$), which causes common rotation of the PSFs. Therefore, if only the image corresponding to one polarization component were used, it would be nearly impossible to distinguish height from orientation, but imaging separately both circular components fully decouples $z$ and $\xi$. A rotation of the two PSFs in opposite directions also occurs when a dipole oriented in the $z$ direction changes height. 
To provide intuition for this behavior, the bottom row of Fig.~\ref{PSFcomps} shows $\partial_z{\cal I}_n^{(p)}$, namely the derivative with respect to $z$ of each of the basis elements at the plane $z=0$. 
The similarity of $\partial_z{\cal I}_{1}^{(p)}$ and $\partial_z{\cal I}_{2}^{(p)}$ with ${\cal I}_{2}^{(p)}$ and ${\cal I}_{1}^{(p)}$, respectively,  explains the fact that both the in-plane rotation and vertical displacement of a horizontal dipole cause rotations; the distinguishability between them arises because $\partial_z{\cal I}_{1}^{(p)}$ and $\partial_z{\cal I}_{2}^{(p)}$ have opposite signs for $p=\mathcal{R}$ and $\mathcal{L}$, while ${\cal I}_{2}^{(p)}$ and ${\cal I}_{1}^{(p)}$ do not, making in-plane orientation and height decoupled in the retrieval process. 

\subsection{Cram\'er-Rao analysis.}
In order to estimate the sensitivity of CHIDO, we use  Cram\'er-Rao (CR) lower bounds\cite{Ober2004,Vella2020} on the uncertainties of the six parameters being measured ($x,y,z,\xi,\theta,\Omega$). These bounds were deduced from a numerical calculation of the inverse of the Fisher Information matrix, in this case of dimension $6\times6$. Each of the six lower bounds depends on all six parameters, as well as on the photon number, the SEO's stress parameter $c$, the pixelation of the PSFs, and the signal-to-background ratio (SBR), namely the ratio of the PSF peak intensity to the uniform illumination background. To reduce the size of the parameter space being explored, we fix $c=1.2\pi$ and assume a pixelation level comparable to that of our experimental implementation. 

Let us start by considering the CR lower bounds for the standard deviations of the directional parameters $\xi$, $\theta$, and $\Omega$. Supplementary Note 3 presents the derivation of the simple  order-of-magnitude estimates of these bounds, based on the near-orthogonality of the PSF components which permits the approximation of the diagonal terms of the Fisher matrix:
\bse
\bea
\sigma_\theta&\approx&\frac2{P_{\rm 3D}\sqrt{6\widetilde{\cal N}}}=\frac{(4\pi)^2}{(8\pi^2-6\pi\Omega+\Omega^2)\sqrt{6\widetilde{\cal N}}},\\
\sigma_\xi&\approx&\frac{\sigma_\theta}{\sin\theta},\\
\sigma_\Omega&=&\frac{\sigma_{P_{\rm 3D}}}{\partial_\Omega P_{\rm 3D}}\approx\frac{1.43}{\partial_\Omega P_{\rm 3D}\sqrt{\widetilde{\cal N}}}=\frac{1.43(4\pi^2)}{(3\pi-\Omega)\sqrt{\widetilde{\cal N}}},
\eea
\label{CRest}
\ese
where $\widetilde{\cal N}={\cal N}/(1+2\,{\rm SBR}^{-1})$, with ${\cal N}$ being the number of signal photons. Here, $\sigma_\xi$ and $\sigma_\theta$ are given in radians and $\sigma_\Omega$ is given in steradians. 
The simple dependence of these bounds on $\theta$ and $P_{\rm 3D}$ suggests a definition for a global measure of directional/wobbling precision as $\sigma_{\rm Dir}=P_{\rm 3D}^2\sin\theta\,\sigma_{P_{\rm 3D}}\sigma_\theta\sigma_\xi\approx\widetilde{\cal N}^{-3/2}$, which depends solely on the number of photons and SBR values. Interestingly, if we were to consider a spherical space where $P_{\rm 3D}$ is the radial variable and $\xi$ and $\theta$ are the azimuthal and polar angles, this measure would correspond to the volume element in this space implied by the CR bounds.
The value of $\widetilde{\cal N}^{3/2}\sigma_{\rm Dir}$ was calculated numerically for 10000 randomly selected cases with inverse SBR between 0 and 3, heights between $-200$ nm and $200$ nm, and the orientational parameters covering uniformly the sphere  with coordinates $(P_{\rm 3D},\xi,\theta)$;  as shown in Supplemental Fig.~3(b), the resulting distribution indeed peaks near unity.

\begin{figure}
\centering
\includegraphics[scale=.6]{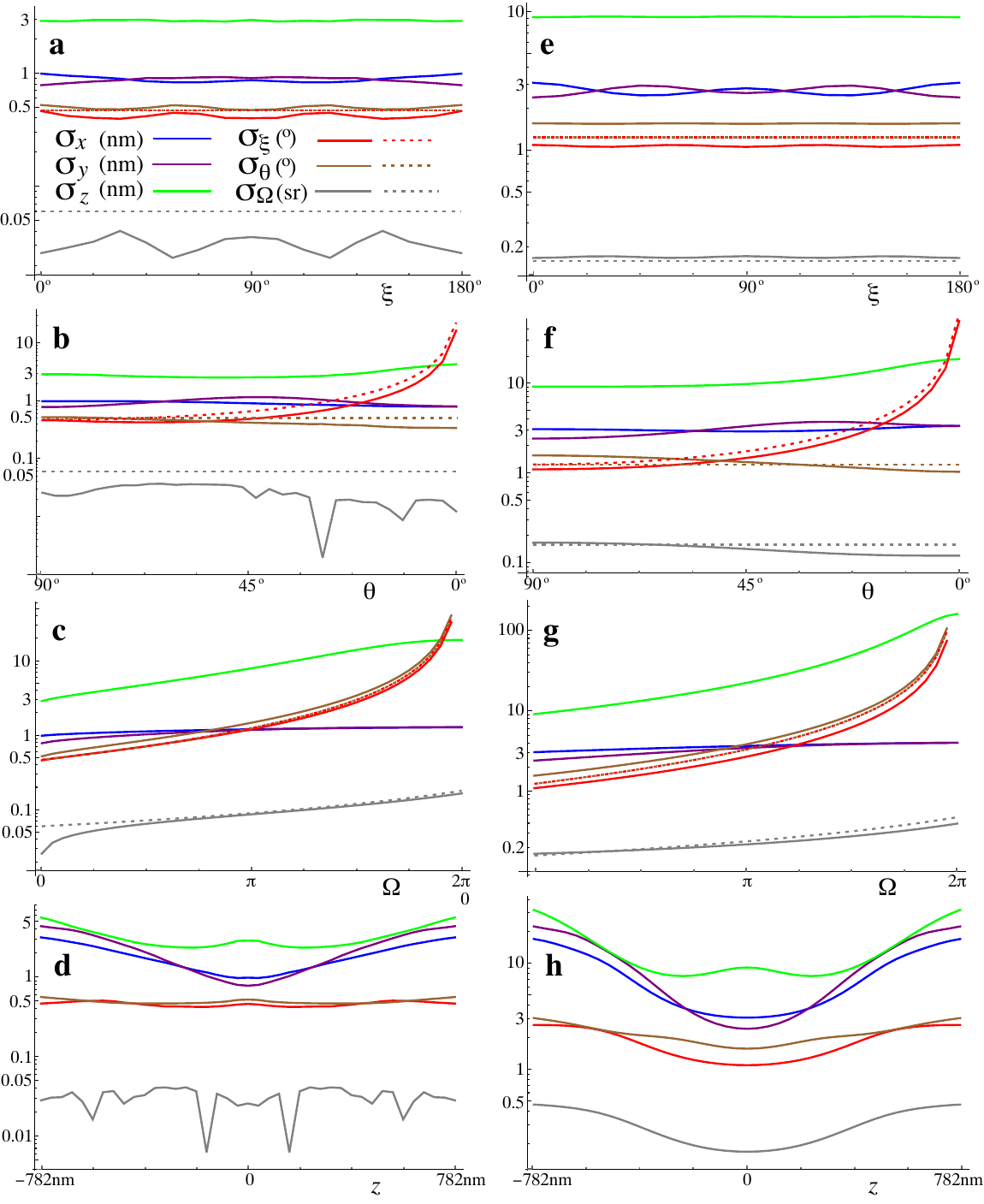}
\caption{
Cram\'er-Rao lower bounds for the six measured parameters. These plots assume $10000$ signal photons over both channels, for (a-d) no background photons, and (e-h) a SBR of $1/3$, corresponding to a background of about 250 photons per pixel. The parameters are: (a,e) $x=y=z=0$, $\theta=90^\circ$, $\Omega=0$ and varying $\xi$; (b,f) $x=y=z=0$, $\xi=0^\circ$, $\Omega=0$ and varying $\theta$; (c,g) $x=y=z=0$, $\xi=0^\circ$, $\theta=90^\circ$, and varying $\Omega$; and (d,g) $x=y=0$, $\xi=0^\circ$, $\theta=90^\circ$, $\Omega=0$ and varying $z$. The units for each curve are indicated in the legend in (a). The dashed lines indicate the simple estimates in Eqs.~(\ref{CRest}).
}
\label{figCR}
\end{figure}
Figure~\ref{figCR} shows comparisons of the simple estimates in Eqs.~(\ref{CRest}) (dotted lines) to numerically-calculated values (solid lines) of the CR lower bounds for several values of the parameters, both in the absence of background photons (a-d), and for a SBR of $1/3$ representative of the single molecule measurements presented later (e-h). The agreement between the approximate expressions and the more rigorous theoretical calculations is very good as expected. The figure also shows the CR lower bounds for the three spatial parameters. For the sake of illustration, these results assume a total of 10000 signal photons over the two detection channels; the obtained levels of error scale as the inverse of the square root of the photon count.
Figures~\ref{figCR}(a,e) show the variations with $\xi$ of the six CR lower bounds for a non-wobbling fluorophore within the plane $z=0$ with in-plane orientation ($\theta=\pi/2$). 
Note that, indeed, the dependence of the lower bounds on the in-plane orientation of the fluorophore is not very significant. 
The variation of the CR bounds as the off-plane orientation changes is illustrated in Figs.~\ref{figCR}(b,f) (assuming $\xi=0$), showing that this change of orientation makes the uncertainty in $z$ first decrease slightly and then increase by less than a factor of two. 
The uncertainty in $\xi$ grows
as the inverse of $\sin\theta$, as expected.
Figures~\ref{figCR}(c,g) show that even moderate amounts of wobble have an adverse effect on the CR bounds for height and direction. These lower bounds are indeed roughly multiplied by 3 when $\Omega$ reaches $\pi$. Finally, given that the PSFs occupy a sufficiently large number of pixels, the CR bounds depend very weakly on changes in $x$ and $y$. The dependence on $z$ is more significant, as can be seen in Figs.~\ref{figCR}(d,h), which shows that 3D spatial localization is affected given the expansion of the PSFs with defocusing; 
the different behavior in $x$ and $y$ is due to the chosen molecule orientation ($\xi=0$). On the other hand, the effect of $z$ over the level of precision of the determination of direction is lower, particularly for low background.  

The estimates just shown, as well as other simulations we performed, indicate that when a few thousand photons are measured, one can expect a precision in transverse position of a few nanometers, and an uncertainty in $z$ about three or four times larger. The corresponding precision in the determination of orientation angles is of a few degrees, and for wobble it is on the order of tenths to hundredths of sterradians. 
These levels of precision are comparable or superior to those of other approaches restricted to the estimation of a subset of the parameters, whether they are based on engineering the PSF\cite{Agrawal,Backer2013,Zhang2018} or on observing the natural change in shape of un-engineered PSFs\cite{Aguet2009,Ding2020}. 
As shown in Supplementary Note 2, these estimations are not only precise but they also involve relatively low levels of coupling between parameters. 
These results were validated by performing Monte Carlo simulations on randomly-generated data with similar signal and background levels (Supplementary Note 3). The retrievals were based on the maximization of the normalized correlation with the model PSFs, which was chosen due to its simplicity and speed. The resulting standard deviations were found to be reasonably close to the CR lower bounds (about 2 to 5 times larger, as can be seen in Supplementary Figure 6), despite the simplicity of the retrieval method. Importantly, no systematic bias in the retrieved parameters was found, even in the presence of background, except near the endpoints of the allowed range of values of $\Omega$, which is natural given the finite, non-periodic nature of this parameter (Supplementary Figures 6 and 7). 

Finally, to evaluate the robustness of the method with respect to image aberrations, the CR calculations were repeated assuming that the system presents one wave of spherical aberration. While this aberration does change the shape of the PSFs, the CR bounds remain largely the same, as shown in Supplementary Figure 5. Note, however, that the presence of aberrations, if not accounted for in the model used for the retrieval of the parameters, will likely introduce (non-uniform) bias in the results, as would be the case for any other super-resolution-based method.


\subsection{Proof-of-principle measurements of height and in-plane orientation with fluorescent beads.}
\begin{figure}
\centering
\includegraphics[scale=0.6]{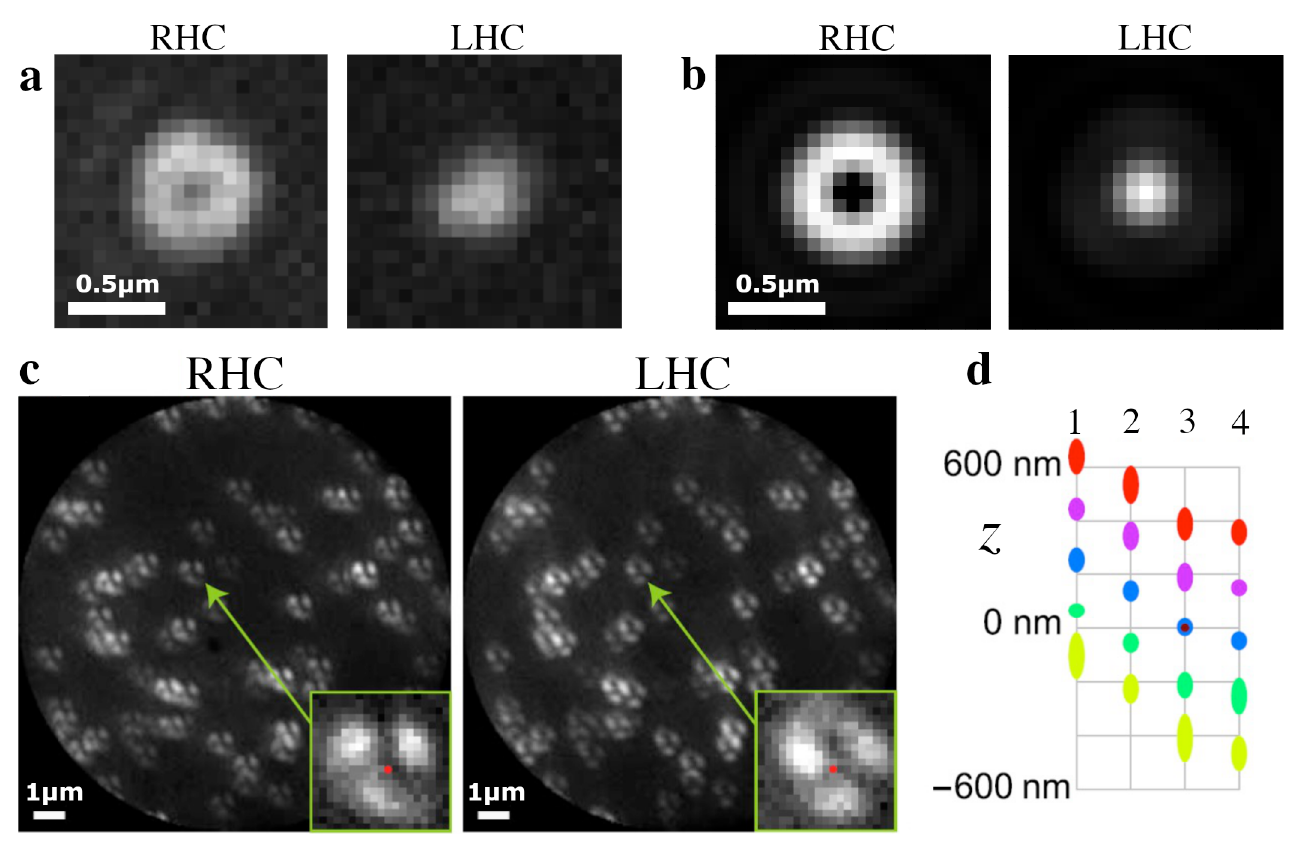}
\caption{Experimental PSFs obtained from nanobeads under circular polarization and simulated $z$ oriented dipole. (a) Measured and (b) simulated images for two nanobeads followed by a circular polarizer. (c) Image pair for a group of nanobeads with a S-waveplate inserted at the pupil plane for simulating emitters oriented in the $z$ direction. For these images the integration time is typically 1 s (camera gain 300). The insets show zooms of the PSFs of a particular bead, where the red dot indicates the retrieved $(x,y)$ coordinates. (d) Estimation of $z$ (average - center of the ellipse, and standard deviation - height of the ellipse) for the five defocused measurements of the four sets of measurements. The brown dot at the central position for set 3 indicates that it corresponds to the images shown in part (c). All retrieved data are depicted in Supplementary Movie 2, including standard deviations for each image.}
\label{zfigs}
\end{figure}

As a first proof of concept and characterization of the method, 
we used fluorescence nanobeads immobilized in a mounting medium (see Methods), together with chosen polarizing elements prior to the SEO
to simulate molecules with known orientations. 
The optical set-up for CHIDO is displayed in Fig.~\ref{setup}(b) (see Methods). A 488 nm continuous laser is used for wide-field illumination of the sample via a high numerical aperture objective (NA = 1.45 oil immersion). The fluorescence ($\lambda=520$ nm) is imaged onto an emCCD camera after passing through the SEO placed at the imaged back focal plane of the objective. 
Importantly, nanobeads are also used to fine-tune the alignment of the SEO when used under circular polarization (see Methods). In such situation, we measured complementary rotationally symmetric PSF shapes in RHC and LHC channels (Fig.~\ref{zfigs}(a)), which are close to what is expected from theory (Fig.~\ref{zfigs}(b)).

Following alignment, measurements were taken for several sets of nanobeads (corresponding to different regions of the same sample) with polarization filters to simulate different fluorophore orientations. For each, measurements were taken at five defocus distances, at separations of 200 nm (see Methods). Rather than using a theoretical model, we chose the PSFs for one bead and used them to construct the PSF model used to extract the parameters for all the beads. The dependence in $z$ of this PSF model was approximated by fitting the measured PSFs of the reference bead with a 
polynomial expression in $z$ of the form
\be
{\cal I}_n^{(p)}(\brho,z)\approx\sum_{m=0}^Mz^m{\cal I}_{n,m}^{(p)}(\brho).
\label{ztothen}
\ee
For simplicity, we used an expansion up to $M=2$, which is sufficient to fit the PSFs at five heights fairly well, but does introduce some systematic errors that limit the range in $z$ over which the retrieval is valid. 
Details of this simple approximate approach for the estimation of $z$, its limitations and ways to improve its range of validity are discussed in Supplementary Note 4. Note that this method can be used as a starting point for a more rigorous parameter retrieval approach using the maximization of the likelihood function or the minimization of rms error. Finally, note that the retrieved direction parameters for polarized nanobeads do not include $\Omega$, since its value is expected to be close to $0$, a region with known systematic bias (see Supplementary Figures 6 and 7). Polarized nanobeads are a good framework to evaluate the robustness of the method for fixed dipoles.

We first investigated the case of emitters oriented in the axial $z$ direction ($\theta=0^\circ$), whose polarization distribution at the pupil plane is radial. To simulate this situation, we inserted a radial polarization converter (Altechna, S-waveplate) before the SEO. This experimental simulation could be made more accurate by also introducing an amplitude filter that simulates the correct radial dependence (approximately linear rather than constant). However, numerical simulations show that the difference in the resulting PSFs is not too significant. Images of four different sets of nanobeads were measured. A typical image taken at the central defocus position is shown in Fig.~\ref{zfigs}(c). 
Using the references constructed from the PSFs of one bead from one of the sets, the transverse and axial positions of the nanobeads for all four sets were detected. Some of the results were discarded due to low confidence (calculated as the normalized correlation of the measured PSFs with those of the model evaluated at the estimated values of the parameter), caused either by low signal levels, overlapping PSFs, or PSFs clipped at the edge of the field of view. The resulting number of nanobeads used for retrieval in set 1 was about 21 on average, while for the remaining sets it was about 35. The average and standard deviations of the retrieved heights for each of the measurements are shown in Fig.~\ref{zfigs}(d). For the four sets, the average estimated heights are separated by approximately 200 nm as expected. This result used a correction in which systematic errors were largely removed by replacing $z$ with an appropriate monotonic function of $z$. (The results without the corrections are shown in Supplementary Fig.~9) Note that from the retrieved 3D positions over the four sets, it was observed that the plane containing the nanobeads was tilted by about a quarter of a degree. More details about the retrieved data from each set are shown in Supplementary Movie 2.

The number of photons detected for each nanobead was of the order of a hundred thousand, but with significant variations as can be appreciated from Fig.~\ref{zfigs}(c) and Supplementary Movie 2. Similarly, the SBR varied from about $1/2$ to about $5$. For each measurement, the spread in the estimates of $z$ was about 25 to 40 nm near $z=0$ (depending on the set), growing to about 60 nm for $z\approx\pm400$ nm (See Supplementary Fig.~10(a)). These spreads are due in part to the possible non-uniformity of the substrate's shape, or to variations in the size of nanobeads (nominally of 100nm). The effect of some of these systematic sources of errors can be removed by considering not the standard deviation of the estimated separations, but the differences of the height estimates at two consecutive heights for each bead, which brings the spreads down slightly, so that for some sets the spread near $z=0$ is of about 20 nm (Supplementary Fig.~10(a)). This is still significantly larger than the CRB predictions from the previous section, which for ${\cal N}=50000$ and SBR = 3 predict $\sigma_z\approx3$ nm. Note, however, that part of the discrepancy emerges from the fact that the fluorescent beads have a size (100 nm) that is not negligible compared to the scale of the PSFs, and their extension both in the transverse and longitudinal directions has the effect of an appreciable blurring of the PSFs. By using the experimentally-obtained PSF model and assuming ${\cal N}=50000$ and SBR = 3 we find instead $\sigma_z\approx4.5$ nm near $z=0$ and 6 to 7 nm for $z=\pm400$ nm, which is only off by about a factor of 4 from the measured standard deviations. This remaining factor is probably due to imperfections in the model, which was based on a single bead. For example, there could be small amounts of field-dependent aberrations over the measured field of view which would deform the PSFs, therefore producing errors in the retrieval.

\begin{figure}
\centering
\includegraphics[scale=.8]{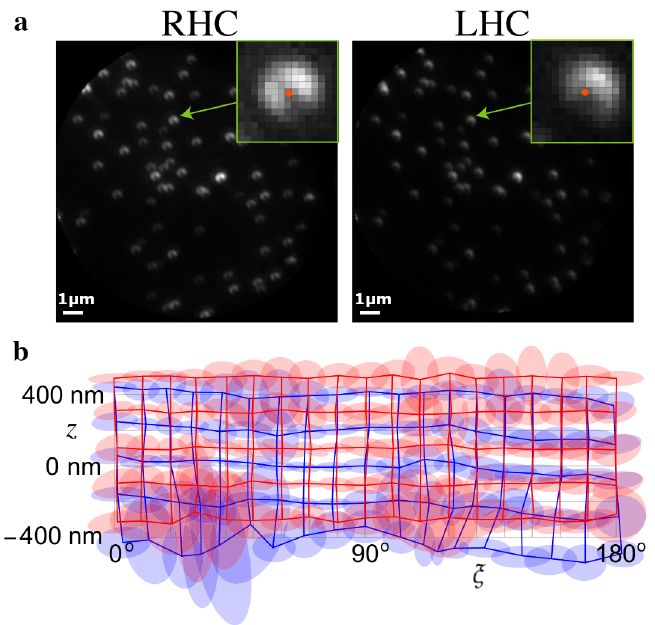}
\caption{Experimental PSFs obtained from nanobeads under linear polarization for simulated oriented dipoles in the $xy$-plane. (a) Images for the two polarization components for set 2 at $\xi=64^\circ$ and at the central defocus position. The insets show zooms of the PSFs of a particular bead, where the red dot indicates the retrieved $(x,y)$ coordinates. (Note that the retrieved coordinates are not at the centroid of the measured PSFs.) For these images the integration time was 1 s (camera gain 300). (b) The intersection points of the blue and red grids indicate the averages of the retrieved heights and orientation angles for each measurement, for sets 1 and 2, respectively, and the ellipses centered at each intersection indicate the corresponding standard deviations. 
A shift of $16^\circ$ was applied to the $\xi$-axis so that the retrieved angles fall in the range $[0^\circ,180^\circ]$ for ease of interpretation (the polarization was rotated from $-16^\circ$ to $164^\circ$).
The full set of data including standard deviations is shown in Supplementary Movie 3 for set 1 and Supplementary Movie 4 for set 2. 
}
\label{linfigs}
\end{figure}
We then simulated emitters with different orientations within the $xy$-plane (i.e. for $\theta=90^\circ$ and varying $\xi$) by replacing the S-waveplate with a linear polarizer prior to the SEO (see Methods). Images were taken for two sets of nanobeads corresponding to two regions of a sample, each at five defocus heights in steps of 200 nm, and for several orientations of the polarizer in steps of $10^\circ$ over a range of $180^\circ$. One of these measurements is shown in Fig.~\ref{linfigs}(a). Again, the measured PSFs from a single bead from one of the sets were selected to generate the PSF model used in the parameter retrieval for the others. Once more, a threshold in the level of confidence of the fit was applied to eliminate errors from overlapping/clipped PSFs and low signals, yielding results for about 30 nanobeads in set 1 and 36 in set 2. The insets in  Fig.~\ref{linfigs}(a) show the retrieved $(x,y)$ position of a specific nanobead. The retrieved heights and orientations and their standard deviations for the two sets are shown in Fig.~\ref{linfigs}(b), whose data are fully displayed in Supplementary Movie 3 and Supplementary Movie 4. An average defocus shift of about 100 nm was found between the two sets. We can also appreciate from the measurements that there was a relative drift in $z$ between both sets of about 100 nm over the time of data collection (over 30 minutes for each). Finally, it can be seen that the large standard deviations for some heights and directions in  Fig.~\ref{linfigs}(b) are caused mostly by a few outliers not filtered out by the confidence threshold, corresponding to PSFs with low intensity, with overlaps, or clipped by the edge of the field of view. In general, we can see that the use of a quadratic approximation for $z$ gives rise to a magnification of the errors at the edges of the interval. The standard deviation in the estimate of $z$ varies greatly from image to image,  with an average of about 39 nm. The corresponding standard deviation of the estimates of each bead's step sizes in $z$ are about 34 nm. The CR lower bound using the PSF model obtained from the beads is $\sigma_z\approx7$ nm, which is about a factor of five smaller (See Supplementary Fig.~10(b)), the reasons for this factor being probably the same as those for the measurements using the S-waveplate. 


\subsection{Single molecules and super-resolution imaging.}
We then applied CHIDO to super-resolution orientational imaging, using fluorophores appropriate for Stochastic Optical Reconstruction Microscopy (STORM)\cite{Dempsey2011}. In order to evaluate the capacity of CHIDO to retrieve both 3D orientations and 3D positions of single molecules, we first imaged Alexa Fluor 488 fluorescent molecules (AF488) sparsely attached to in-vitro reconstructed F-actin single filaments via phalloidin (see Methods). These molecules are known to keep an average orientation along the actin filament, with a non-negligible wobbling extent\cite{ValadesCruz2016}.

\begin{figure}
\centering
\includegraphics[width=0.35\linewidth]{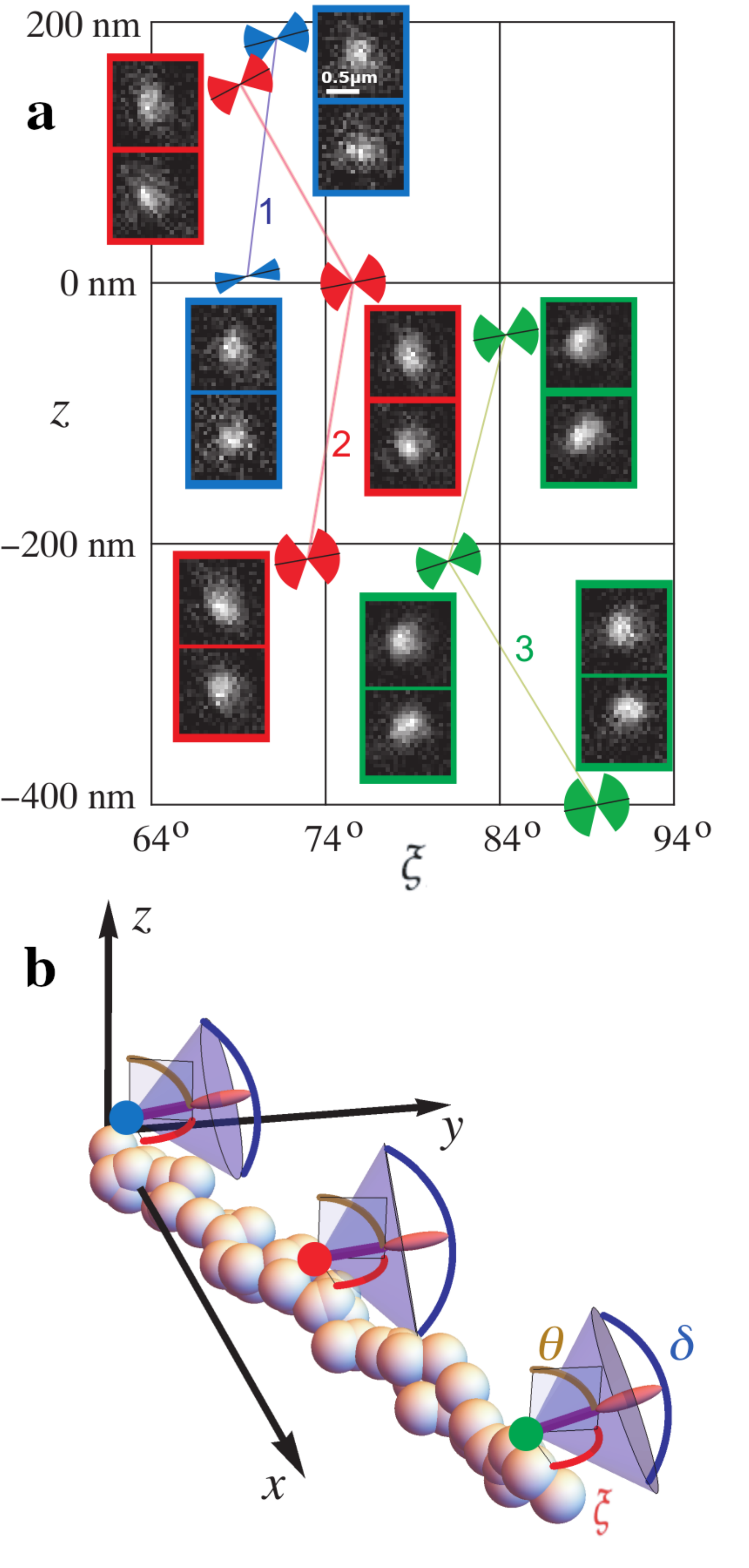}
\caption{CHIDO imaging of three Alexa Fluor 488 single molecules sparsely attached to a F-actin filament. (a) The insets show zooms of the images of the RHC/LHC components on the top/bottom for three different $z$ defocus (only high confidence retrieval are shown). These images are labeled by the color of the frame: blue for molecule 1, red for molecule 2, and green for molecule 3. 
The estimated in-plane angle $\xi$ and height $z$ are represented as coordinates in this plot, where the PSFs used to retrieve the corresponding values are placed closest to the corresponding point. The out-of-plane angle $\theta$ and wobble angle $\delta$ are represented by the orientation and angular extent of the sketched cones. (b) 3D representation of the orientation and wobbling retrieval for defocus $z$ close to 0. Numerical values are given in Table~\ref{moleculedata}.
}
\label{molecules}
\end{figure}
For the first set of measurements, we retrieved the localization and orientation of the fluorophores by using a PSF basis set constructed from a combination of theoretical calculations and the reference nanobead measurements using the s-wave plate and linear polarizers. As explained in Supplementary Note 4, there are several potential problems with the resulting PSF model, which can arise from the combination of the models for in-plane and out-of-plane orientations, which used different beads whose relative 3D positions and brightnesses are not know accurately. Also, as mentioned before, beads have extensions that are much larger than that of single fluorophores, so the resulting PSF model is smoother. Figure~\ref{molecules} shows the results obtained for three molecules positioned along an F-actin filament. For each, three sets of images were taken at defocus separations of 200~nm. Isolated pairs of PSFs around the retrieved positions (after subtraction of the average background) are shown in the insets of Fig.~\ref{molecules}(a). The retrieved 3D positions, orientations and wobble angles of these molecules are presented in Figs.~\ref{molecules}(a,b) and in Table~\ref{moleculedata}, with the exception of that for the top position for molecule 1, which fell outside the range of validity of the model generated from the nanobeads reference measurements. The range of in-plane orientations $\xi$ measured for the three molecules is restricted to a $30^\circ$ interval, which is expected from their attachment to a single oriented filament. The off-plane angle $\theta$ and wobble angle $\delta=2{\rm arccos}(1-\Omega/2\pi)$ are also consistent with expectations: polarized measurements performed in 2D have shown fluctuations within an extent $\delta$ of about $90^\circ$, with a tilt angle with respect to the fiber that can reach $20^\circ$\cite{ValadesCruz2016}. In the course of the measurements at different $z$ positions, the retrieved transverse positions (with respect to the center of the selected insets) present an uncertainty (namely the averaged standard deviation for each molecule, weighted by the number of measurements) of about 13 nm in both $x$ and $y$, and a corresponding directional uncertainty of about five degrees. The estimated defocus spacings average to 198 nm but have an uncertainty of almost 50 nm. The uncertainty in the wobble solid angle is just below $0.9$. Given the long integration times, a total of about 40000 photons were detected for each molecule, with a SBR of about 1/3, so that according to the CR analysis the uncertainties for the three spatial coordinates are about six times larger than the CR lower bound ($\sigma_{x,y}\approx 2.3$ nm and $\sigma_z\approx 7.5$ nm), and similarly for the directional parameters  ($\sigma_\theta\approx 0.8^\circ$ and $\sigma_\Omega\approx 0.13$). Recall that the PSF basis used for the retrieval was constructed from measurements for 100 nm beads, which do lead to slightly larger CR lower bounds than point dipoles. 



\begin{figure}
\centering
\includegraphics[width=.75\linewidth]{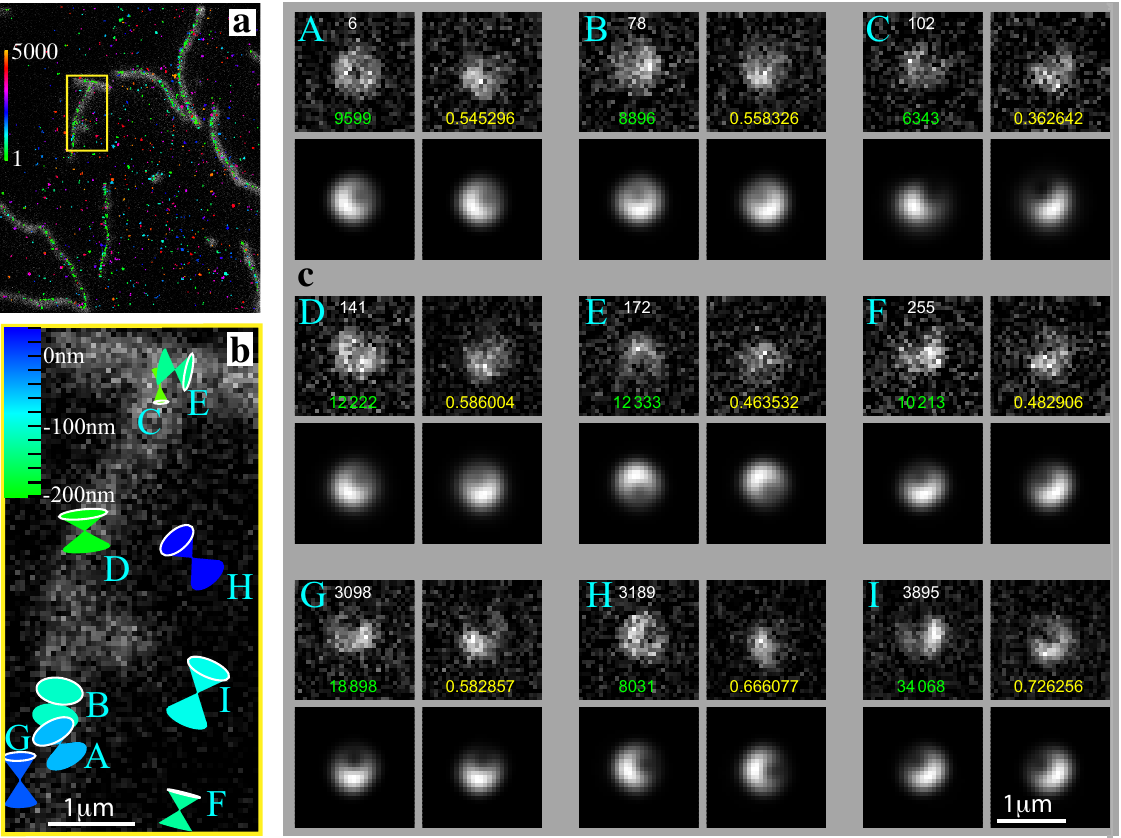}
\caption{
CHIDO analysis using STORM imaging on single F-actin filaments labeled with AF488 phalloidin conjugates. (a) Detected single molecules in a STORM stack of about 5000 images, where the color indicates the stack number, shown over the low resolution image of single filaments (gray background). (b) For the region within the yellow rectangle in (a), detected single molecules exhibiting a sufficiently high retrieval confidence level. The symbols correspond to 2D projection of 3D cone pairs whose vertex encodes $x,y$, whose color represents height according to the scale, whose orientation represents that of the fluorophore, whose solid angle represents $\Omega$, and whose size reflects the correlation between the detected and model PSFs. (c) PSF pairs for each of the detected molecules in (b), where the top row shows the two detected PSFs and the bottom row the PSFs from the model evaluated at the retrieved parammeters. For each part, the white numbers indicate the frame at which the PSF appeared, the green number gives the estimated number of signal photons, and the yellow number is the correlation between the detected and model PSFs.
}
\label{STORM}
\end{figure}
Finally, we applied CHIDO to samples imaged in STORM conditions, e.g. 
single F-actin filaments labelled densely with AF488, within a buffer appropriate for on-off blinking conditions in order to localize individual emitters (see Methods and Supplementary Movie 5). The integration time was lowered to 200 ms, leading to significantly more challenging signal conditions than the single molecule measurements described above. 
Estimates of 3D localization and orientation were performed on each detected single molecule by fitting the measured PSFs to theoretical model PSFs. We used a model based on theory in order to adapt to the slightly higher $c$ value used for this experiment (see Methods), and to overcome the limitations of reference PSFs generated with extended beads. Note that aberrations and misalignments in the system not accounted for by the model may induce some bias in the determined parameters, although not affecting precision. Figure~\ref{STORM}(a) shows the rough positions of all detected molecules (color-coded by frame number) in a stack of about 5000 STORM image frames, on which single filaments are also identified by their low-resolution image. Note that after about 1000 frames, blinking seems to be dominated by molecules that are not attached to the filaments. 
In the collection of molecules measured, SBR values typically range from 0.2 to 1.2, with a large population around 0.3.
Figure~\ref{STORM}(b) depicts the resulting retrieved parameters for molecules within the highlighted section of the image, where we consider only molecules for which the normalized correlations of the measured PSFs (after background subtraction) to the model is above a threshold of 0.35. These molecules exhibit 3D and wobbling information that are in agreement with expected values. Notice that the range of retrieved heights is large, possibly due in part to inaccuracy introduced by the theoretical model used.
Nevertheless, we can observe that the results are consistent with filaments 
laying on top of each other, as shown by the two molecules (C and E) near the junction of the two filaments, whose orientations are nearly perpendicular and whose heights are notably different. The measured PSFs for all these molecules are compared to those of the theoretical model evaluated at the retrieved parameters in Figure~\ref{STORM}(c), which also shows the frame number, estimated number of photons, and normalized correlation between measured and model PSFs.
The relatively large photon levels for some of the molecules depicted in Fig.~\ref{STORM}(c) is due to the sum performed when a single molecule appears 
in several consecutive frames in the STORM images stack. Several molecules of Fig.~\ref{STORM}(c) depict experimental PSFs that are slighlty visually different than the retrieved PSFs, which we attribute to the presence of slight aberrations and/or misalignments (e.g. decentering of the SEO) in the optical setup, that are not included in the model. Such small imperfections can be the source of some degree of bias, that could be accounted for in a more complete model development that will be the focus of future efforts. Note that, according to the CR bounds, the highest precision of the retrieved positions is expected to be of the order of a nanometer and of the direction of about a degree. Furthermore, the precision in 
the solid angle wobble is of hundredths to about a tenth of a steradian. These expected levels of precision show that, when a proper reference model is used, CHIDO is applicable to STORM imaging conditions.

\begin{table}
\centering
\caption{\bf Retrieved positions, orientations and wobble angle for the fluorescent  molecules in Fig.~\ref{molecules}(a,b). The standard deviations on the last column are weighted averages of the standard deviations for each molecule, with the exception for that in $z$, which is the weighted average of the standard deviations for each molecule of the increments in height.}
\begingroup
\setlength{\tabcolsep}{3.7pt} 
\renewcommand{\arraystretch}{1} 
\begin{tabular}{c|c c|c c c|c c c|c}
&\multicolumn{2}{c}{Molecule 1}&\multicolumn{3}{c}{Molecule 2}&\multicolumn{3}{c}{Molecule 3}&
\\
&$z_1$&$z_2$
&$z_1$&$z_2$&$z_3$&$z_1$&$z_2$&$z_3$&St. dev.  \\
\hline
$x$ (nm) &$-108$&$-98$
&$-72$&$-93$&$-109$&$-132$&$-118$&$-114$&$12.3$
\\
$y$ (nm)&$110$&$137$
&$62$&$76$&$38$&$23$&$19$&$15$&$13.5$
\\
$z$ (nm)&$-2$&$221$
&$-218$&$8$&$121$&$-499$&$-268$&$-75$&$49.6$
\\
$\xi$&$66^\circ$&$71^\circ$
&$74^\circ$&$77^\circ$&$72^\circ$&$89^\circ$&$83^\circ$&$88^\circ$&$3.03^\circ$
\\
$\theta$ &$77^\circ$&$75^\circ$
&$80^\circ$&$79^\circ$&$62^\circ$&$79^\circ$&$72^\circ$&$80^\circ$&$5.78^\circ$
\\
$\Omega$&$0.4$&$1.5$&$3.2$&$2.3$&$1.7$&$3.7$&$1.8$&$2.2$&$0.88$
\\
\end{tabular}
\endgroup
\label{moleculedata}
\end{table}

\section*{Discussion}
A method, CHIDO, was proposed that allows the measurement of the 3D position, averaged 3D orientation and wobbling of isolated fluorophores, readily applicable for super-resolution orientational imaging. The key elements of this method are a specific spatially-varying birefringent mask, the SEO, inserted at the pupil plane, and the subsequent separation of the two circular components to form separate images. The use of both images is shown to be of central importance for decoupling the estimations of in-plane orientation and axial position $z$. Despite the large amount of information encoded in the shape of the PSFs, these have dimensions that are only about twice as large as the corresponding diffraction-limited PSFs, making this approach suitable for measurements with relatively high densities of molecules since their PSFs would not overlap significantly if their separations are about a micron or even less. The compactness of the PSFs also helps maintaining a higher SBR for the measured PSFs. Importantly, CHIDO is 
satisfactorily achromatic over the detected spectral range of fluorescence. 


The retrieval of the parameters requires a reliable model for the PSFs. For the proof-of-concept measurements that emulate molecule orientation using beads and appropriate polarizers, the models were obtained by using the measurements from one of the beads and a polynomial interpolation in $z$. 
These references were also applied to the first set of measurements for single molecules. However, the construction of the PSF basis was based on reference measurements for orientations within the $xy$-plane and normal to it, and a complete set of PSFs also requires measurements at intermediate off-plane angles (e.g. $\theta=45^\circ$), which are more difficult to simulate experimentally (one imperfect option being an off-center S-waveplate). This incompleteness was addressed by using a mixture of theoretical simulations and experimental data. 
For the STORM-like measurements, however, a slightly higher value of $c$ was used, which produced PSFs with finer features. We found that for such PSFs the blurring resulting from using 100 nm reference beads was more critical, so instead we used a theoretically generated model. 
However, using a theoretical model that does not exactly account for specific characteristics of the imaging system might introduce systematic errors. In general, one of the main lines of research that we will explore in the future is the obtention of a reliable PSF model through a combination of theoretical and experimental approaches. These could include the use of smaller fluorescent beads combined with polarizers, appropriate deconvolution methods, and phase retrieval techniques that use measurements at several heights. These models will also seek to characterize the effects of field-dependent aberrations, so that transverse position can be incorporated into the model beyond a simple (pixelized) translation.


A future direction to be explored is to use CHIDO not only to estimate the amount of wobbling of the molecules but also the asymmetry of this wobbling\cite{Beausang2013,Backer2015}. As discussed in Supplementary Note 1, the $3\times3$ correlation matrix in Eq.~(\ref{Gamma}) encodes information about the correlation of all field components\cite{Backer2015}, which in the context of vector coherence corresponds to the shape of an ellipsoid that characterizes the oscillations of the field\cite{Mark}.  
Within the current context, this translates into the capacity of estimating not only a solid angle but, say, two angles of oscillation for the molecule supplemented by an angle of orientation of this elliptical cross-section of the cone. We expect that with refinements of the system, and more importantly of the PSF basis, it will be possible to recover useful information about these extra paramenters for single molecules, which can then be compared with computational models for the molecular motion. Finally, while CHIDO was restricted here to non-overlapping PSFs, new fitting procedures could be developed to adapt the method to samples with higher densities, following recent work in the field\cite{Holden2011, Huang2011, Zhu2012, Mailfert2018, Barsic2015, Mazidi2019}.

With these possibilities, other applications for CHIDO can emerge in addition to imaging the 3D position and orientation of fluorophores. For example, this method could be used to probe the $3\times3$ correlation matrix at several points of a strongly nonparaxial field, such as a focused field or an evanescent wave. This would require the use of one or an array of sub-wavelength scatterers such as gold nanoparticles\cite{Lindfors,Lukas}.

\begin{methods}
\subsection{Optical setup.}
The sample is excited by a laser (Coherent, Obis 488LS-20 for reference beads and single molecule measurements; Coherent, Sapphire 488LP-200 for STORM measurements) in a wide-field or TIRF illumination configuration (Fig. \ref{setup}(b)), and is held on a piezo nanopositioner (Mad City Labs Inc., Nano-Z200) to perform stacks along the $z$ axis with nanometric precision. Fluorescence light emitted by the sample is then collected by a $\times 60$, NA $1.45$ oil immersion objective (Nikon, CFI Apo TIRF). A dichroic mirror (Semrock, DI02-R488) and a fluorescence filter (Semrock, 525/40) are used to select the emitted fluorescence and send it to the detection path. To adjust the field of view, a diaphragm is placed in a plane conjugate to the image. 
All the lenses are achromatic doublets: L$_1$ (125mm) and L$_3$ (500mm) are in a 4f configuration enabling us to locate the SEO in the back-focal plane;
L$_2$ (400 mm) is used for back-focal plane imaging. To simulate emitters with different in-plane orientations, we put prior to the SEO a linear polarizer (Thorlabs, LPVISE100-A) mounted on a motorized rotation stage (Newport, PR50CC). To compensate unwanted polarization distortions introduced by the first dichroic mirror, we used another identical dichroic mirror (Semrock, DI02-R488), aligned along the plane where \textit{s} and \textit{p} polarization components of the incident fluorescence are inverted with respect to the incidence on the first dichroic mirror. Finally, the image is split into LHC and RHC polarization components by using a quarter-wave plate followed by a quartz Wollaston polarizing $2.2^\circ$ beamsplitter (Edmunds, 68-820), and each of these components is projected onto a different region of an emCCD camera 
(Andor iXon Ultra 897 for beads and single molecule measurements; iXon Ultra 888 for STORM measurements). The total magnification provided by the lenses is $240$, corresponding to a pixel size of $67$ nm on the emCCD for the bead and single molecule measurements, and $54$ nm for the STORM measurements.

\subsection{Stress-engineered Optic.} 
The term stress engineering in optics applies to the design and application of stress birefringence to achieve a desired retardance distribution.   More specifically, the stress engineered optic (SEO) used for this work was one of a collection of SEOs custom fabricated for the T.G. Brown research group at the Institute of Optics, University of Rochester.   Details of the analysis and fabrication of the elements can best be found in the PhD dissertations of Alexis K. Spilman \cite{Spilmanthesis} and Amber M. Beckley \cite{Beckleythesis} and summarized in several related publications \cite{Alexis,Roshita,Brandon}.   To our knowledge, these elements are not yet commercially available but can be readily manufactured by a skilled machine shop. 

The window material can be any transparent material with a nonzero stress optic coefficient; we have used both fused silica and BK7 glass windows.  The material must also be strong enough to withstand approximately 100 MPa of peripheral pressure without fracture.  For the SEO used in this experiment, commercial BK7 windows (10 mm diameter, 8 mm thickness) were first given a fine grind in order to ensure a cylindrical edge.  A metal ring (steel) with inner diameter about 25 microns smaller than the outer diameter of the glass was cut on a lathe.  An end mill was then used to remove material at $0^\circ$, $120^\circ$, and $240^\circ$, leaving three contact points at $60^\circ$, $180^\circ$, and $300^\circ$.   

The assembly is accomplished by heating both the glass and metal to a temperature of $300^\circ$ C; at this temperature, the higher thermal expansion coefficient of the metal allows the insertion of the glass piece into the metal ring.  Upon slow cooling, the ring then compresses the glass, applying force at three small regions around the perimeter in a way that is approximately uniform along the thickness.

\subsection{SEO alignment.}
For the purpose of aligning the SEO and adjusting the parameter $c$, we used a sample of yellow highlighter's fluorescent ink, embedded in a mounting medium (Sigma, Fluoromount). The fluorescence emitted by this sample is used as a bright and homogeneous illumination for the SEO. Circular polarization was produced by placing a linear polarizer and QWP before the SEO.  Also, a lens (L$_2$) was inserted to image the SEO plane, leading to complementary rotationally symmetric intensity patterns whose radial dependence for the two emerging circular components is approximately proportional to $\cos^2(cu/2)$ and $\sin^2(cu/2)$ respectively for $c=\pi$ (Fig.~\ref{setup}(c)). The system's alignment and calibration is then fine-tuned by removing L$_2$ and keeping the polarizer and QWP, to image model nano-emitters under circular polarization conditions. We used for this purpose fluorescent nanobeads of 100 nm in size (yellow-green Carboxylate-Modified FluoSpheres), immobilized on the surface of a poly-L-lysine coated coverslip and covered with a mounting medium (Sigma, Fluoromount). Ideally, the resulting images are complementary, nearly rotationally symmetric PSF shapes, one of them donut-like, the other a bright spot, as shown in Figs.~\ref{zfigs}(a,b)\cite{Roshita}. These shapes are robust under defocus but they are sensitive to polarization distortions, so they can also be useful for calibrating residual undesired birefringence. 
Once this stage of the calibration was complete, the polarizer and QWP prior to the SEO were removed.

\subsection{Single molecule imaging.}
To produce in-vitro reconstituted F-actin filaments, G-actin (AKL99, Cytoskeleton, Inc.) was polymerized at 5 $\mu$M in a polymerization buffer (5 mM Tris-HCl at pH 8.0, 50 mM KCl, 1 mM MgCl2, 0.2 mM Na2ATP at pH 7.0, 1 mM DTT) in presence of 5 $\mu$M phalloidin to stabilize the polymerization. To make the labeling sparse enough to isolate single molecules, we used a ratio of 1:200 phalloidin conjugated to Alexa Fluor 488. The filaments were then diluted to 0.2 $\mu$M, immobilized on the coverslip surface coated with poly-L-lysine and covered with an imaging buffer containing an oxygen scavenging system (5 mM Tris-HCl at pH 8.0, 50 mM KCl, 1 mM MgCl2, 0.2 mM Na2ATP at pH 7.0, 1 mM DTT, 1 mM Trolox, 2 mM PCA, 0.1 $\mu$M PCD). The typical experimental conditions for single molecules imaging were : TIRF illumination, laser power of a few mW (at the objective plane), camera gain 300 and 1s integration time.

\subsection{STORM imaging.}
The F-actin filaments used for STORM imaging were obtained, as for the single molecule images, from G-actin (AKL99, Cytoskeleton, Inc.) polymerized at 5 $\mu$M in a polymerization buffer (5 mM Tris-HCl at pH 8.0, 50 mM KCl, 1 mM MgCl2, 0.2 mM Na2ATP at pH 7.0, 1 mM DTT). To fully label the actin monomers the polimerization was done in the presence of 5 $\mu$M phalloidin conjugated to Alexa Fluor 488. The filaments were then diluted to 0.2 $\mu$M, immobilized on the coverslip surface coated with poly-L-lysine and covered with a STORM buffer (100 mM Tris-HCl at pH 8.0, 10\% glucose, 5 u/ml pyranose oxidase, 400 u/ml catalase, 50 mM $\beta$-MEA, 1 mM ascorbic acid, 1 mM methyl viologen, 2 mM COT).  
Before taking  images, the system was realigned, the value of $c$ was adjusted to $1.2\pi$ (in order to benefit from slightly more complex PSFs) and the SEO was aligned so that one of its stress points pointed in the $y$ direction. The typical experimental conditions were : TIRF illumination, laser power 150 mW, camera gain 300 and 200 ms integration time. For STORM imaging, a stack of 
 5000 
images was used. For each frame, the approximate $x,y$ positions of the fluorophores were detected. Since some fluorophores blinked for longer than the exposure time of one image, a routine was written to sum over all the relevant consecutive frames for each fluorophore in order to reduce SNR. Some fluorophores blinked for up to about ten frames, resulting in photon numbers of up to about 50000. Pairs of arrays of $29\times29$ pixels containing each PSF set were then used to retrieve the parameters.

\end{methods}

\section*{Supplementary Note 1: Theoretical description of the system}
\subsection{Modeling of the optical system.} 
The goal of CHIDO is to image electric dipoles (fluorescing molecules) with a high NA microscope in order to determine their 3D location, orientation and wobbling from the location and shape of their PSFs. 
In this Supplementary Note we describe the forward model used to calculate these PSF shapes. Like for other methods, this model is based on the second moments of the dipole direction from which the orientation-dependent variation of both the far-field\cite{Lieb:04} and the focused light at the detector\cite{Bohmer:03,Aguet2009,Backer2015,HieuThao:20}  can be predicted.

Let the field at the system's pupil plane be denoted by ${\bf E}_{\rm pupil}(\bu)$, where $\bu$ is a dimensionless normalized pupil position with polar coordinates $(u,\varphi)$ so that $u=1$ corresponds to the edge of the pupil. For any orientation of a dipole, ${\bf E}_{\rm pupil}$ can be expressed as a linear superposition of three fields, corresponding to the responses to electric dipoles oriented in the $x,y$ (in-plane) and $z$ (out-of-plane) directions. The field is highly collimated at the pupil plane, so its $z$ component is unimportant. If the dipole is shifted in $z$ from the nominal object plane, the field at the pupil acquires a chirp factor whose phase is proportional to this displacement, while displacements in $x$ and $y$ introduce a linear phase factor. Let us for now ignore these transverse displacements and focus on the axial displacement and the orientation of the dipole. The field at the pupil can then be written as
\be
{\bf E}_{\rm pupil}(\bu)=\sum_{i}E_i{\bf K}_i(\bu)\exp[-\ui k{\rm n}z \gamma(u)],
\ee
where $E_i$ is the $i$th component ($i=x,y,z$) of the field generated by the molecule, ${\bf K}_i(\bu)$ is the amplitude and polarization distribution at the pupil plane generated by each of these components and assumed to differ from zero only for $u\le1$, $k$ is the wavenumber, n is the refractive index of the medium embedding the fluorophores, and for an aplanatic system $\gamma(u)=(1-u^2\sin^2\theta_0)^{1/2}$ with $\theta_0$ being the entrance half-angle of the objective lens. 

If the imaging system is axially symmetric, the functions ${\bf K}_i(\bu)$ must take the general form
\bse\label{second}
\bea
{\bf K}_x(\bu)&=&\hat{\bf x}\,g_0(u)+(\hat{\bf x}\cos2\varphi+\hat{\bf y}\sin2\varphi)g_2(u),\\
{\bf K}_y(\bu)&=&\hat{\bf y}\,g_0(u)+(\hat{\bf x}\sin2\varphi-\hat{\bf y}\cos2\varphi)g_2(u),\\
{\bf K}_z(\bu)&=&(\hat{\bf x}\cos\varphi+\hat{\bf y}\sin\varphi)g_1(u),
\eea
\ese
where, if the system is also aplanatic, the functions $g_n(u)$ are given (to within an unimportant constant factor) by
\bse\label{third}
\bea
g_{0,2}(u)&=&\frac{t_{\rm p}(u)\gamma(u)\pm t_{\rm s}(u)}{2\sqrt{\gamma(u)}},\\
g_1(u)&=&\frac{\sin\theta_0\,u\,t_{\rm p}(u)}{\sqrt{\gamma(u)}},
\eea
\ese
with $t_{\rm p}(u)$ and $t_{\rm s}(u)$ being the transmission coefficients for the radial and azimuthal components, respectively. These coefficients account not only for interfaces inside the microscope, but also for passage from the medium containing the fluorophores to glass to the immersion oil being used.

The basis of this method is to use a stress-engineered optical element (SEO) for wavefront coding at the pupil. This element is described by the Jones matrix $\mathbb{J}(\bu)$ given in Eq.~(1) of the main text. 
After passing through the SEO, the field at the pupil is given by
\be
{\bf E}_{\rm SEO}(\bu)=\mathbb{J}(\bu){\bf E}_{\rm pupil}(\bu).
\ee
The field is then focused to form an image. This focusing corresponds to Fourier transformation:
\bea
\!\!\!\!{\bf E}_{\rm det}(\brho)
&=&\frac{R^2}{\lambda f}\int{\bf E}_{\rm SEO}(\bu)
\exp\left(-\ui \frac{kR\,\bu\cdot\brho}f\right)\ud^2u
={\cal F}{\bf E}_{\rm SEO},
\label{Eq:FourierT}
\eea
where $R$ is the physical radius of the pupil, $f$ is the focal length of the focusing system, and $\brho=(\rho\cos\phi,\rho\sin\phi)$ are the coordinates at the detector plane. In what follows we use the shorthand ${\cal F}$ for the Fourier transformation from the pupil to the detector plane that includes these physical parameters.

\subsection{PSF and generalized Stokes parameters.}
As mentioned in the main body, CHIDO relies on forming separate images for the two circular polarization components emerging from the SEO, by using a combination of a quarter-wave plate and a Wollaston prism. In principle, each of these images would be sufficient to determine the orientation of the emitter if its $z$ coordinate were known. However, using both images allows decoupling the effects of orientation and $z$ displacements on the PSFs, which is a main feature of this technique. It also allows utilizing all photons, leading to more accurate estimates, which is particularly important when photons are scarce. 

To make the following analysis general, we consider separation into any two orthogonal polarization components, represented by the unit vectors $\hat{\bf p}_p$ for $p=1,2$. The corresponding intensity images at the two regions of the CCD are then given by
\be
I^{(p)}(\brho)=\sum_{i,j}\Gamma_{ij}G_i^{(p)*}(\brho)G_j^{(p)} (\brho),
\ee
where $\Gamma_{ij} =\langle E_i^*E_j\rangle$ (in the $x,y,z$ basis) is an element of the correlation matrix $\mathbf{\Gamma}$ of the different components of the (possibly wobbling) dipole source, and
\be
G_i^{(p)} (\brho)={\cal F}\{\hat{\bf p}_p^*\cdot\mathbb{J}(\bu)\cdot{\bf K}_i(\bu)\exp[-\ui kz \gamma(u)]\}.
\label{G1}
\ee

As described in Eq.~(3) of the main text, the intensity corresponding to each polarization can be expressed as a sum of a basis of PSFs ${\cal I}_n^{(p)}$ weighted by the generalized Stokes parameters $S_n$ of the $3\times3$ correlation of the emitter. The expressions for these PSF basis elements are
given by
\bse\label{PSFcomponents}
\bea
{\cal I}_0^{(p)}(\brho)&=&\frac{\left|G_x^{(p)} (\brho)\right|^2+\left|G_y^{(p)}(\brho)\right|^2+\left|G_z^{(p)}(\brho)\right|^2}{\sqrt{3}},\\
{\cal I}_1^{(p)}(\brho)&=&\left|G_x^{(p)} (\brho)\right|^2-\left|G_y^{(p)}(\brho)\right|^2,\\
{\cal I}_2^{(p)}(\brho)&=&2\Re\left[G_x^{(p)*}(\brho)G_y^{(p)}(\brho)\right],\\
{\cal I}_3^{(p)}(\brho)&=&2\Im\left[G_x^{(p)*}(\brho)G_y^{(p)}(\brho)\right],\\
{\cal I}_4^{(p)}(\brho)&=&2\Re\left[G_x^{(p)*}(\brho)G_z^{(p)}(\brho)\right],\\
{\cal I}_5^{(p)}(\brho)&=&2\Im\left[G_x^{(p)*}(\brho)G_z^{(p)}(\brho)\right],\\
{\cal I}_6^{(p)}(\brho)&=&2\Re\left[G_y^{(p)*}(\brho)G_z^{(p)}(\brho)\right],\\
{\cal I}_7^{(p)}(\brho)&=&2\Im\left[G_y^{(p)*}(\brho)G_z^{(p)}(\brho)\right],\\
{\cal I}_8^{(p)}(\brho)&=&\frac{\left|G_x^{(p)}(\brho)\right|^2+\left|G_y^{(p)}(\brho)\right|^2-2\left|G_z^{(p)}(\brho)\right|^2}{\sqrt{3}}.
\eea
\ese
Suppplementary Fig.~1 shows these elementary PSFs and their derivatives in $z$ for $z=0$, where unlike in the main text we show also the ones associated with helicity ($n=3,5,7$), which are not used for dipole orientation.

\begin{figure}[htbp]
\centering
\includegraphics[scale=0.5]{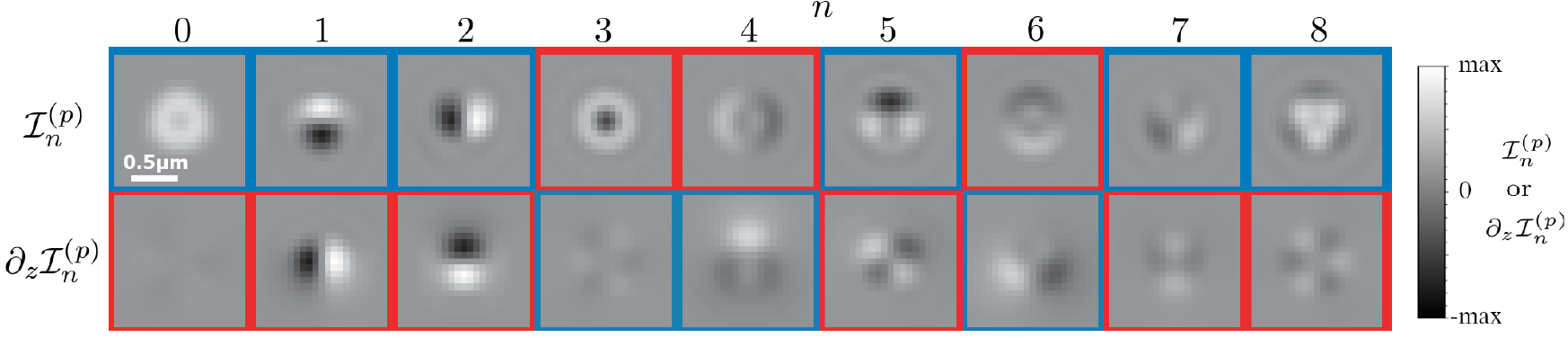}
\caption*{Supplementary Figure 1: Complete set of theoretical PSF components in CHIDO imaging. The figure shows both ${\cal I}_n^{(p)}$ and $\partial_z{\cal I}_n^{(p)}$ for $c=\pi$, $z=0$, and $p=\mathcal{R}$. The corresponding components for $p=\mathcal{L}$ are identical, except that those surrounded by red boxes would have the opposite sign.
Each row is normalized separately as their units are different. }
\label{PSFs}
\end{figure}

\subsection{Chromatic dependence of the PSF.}
Note that the treatment above assumes monochromatic illumination. However, the fluorescent emission from the molecules is not strictly monochromatic but includes a spectral range of about 40 nm around a peak wavelength of 519 nm after passing through a fluorescence filter. The calculation of the PSF components ${\cal I}_n^{(p)}$ then requires the superposition of the corresponding PSFs weighted by the spectrum. It turns out, nevertheless, that the parameter $c$ is approximately inversely proportional to the wavelength because it characterizes a phase retardance. This spectral dependence then balances out that resulting from the presence of the wavenumber in the exponent of the kernel of the Fourier transform in Eq.~(\ref{Eq:FourierT}) for propagating from the pupil to the image plane, making the contributions from each spectral component largely consistent. The resulting PSF components, over a 100 nm spectral expansion, are seen to be very similar to those corresponding to the peak wavelength. 
This is illustrated in Supplementary Fig.~2 for the six relevant PSF components at $z=0$. We can see that even if the fluorescence filter is not used, the chromatically-integrated PSF elements are very similar to those corresponding to the nominal wavelength. Note that this property stays valid for an extent in $z$ of about 500 nm, above which the errors on the PSF can surpass 5\% because the rate of rotation of the PSFs with defocus depends approximately on the ratio $z/\lambda$. A reduction of the detected spectral width down to 40 nm (which is the case in the present experiments) permits to benefit from the achromaticity property of CHIDO even at large shifts in $z$.  

\begin{figure}[htbp]
\centering
\includegraphics[width=0.85\linewidth]{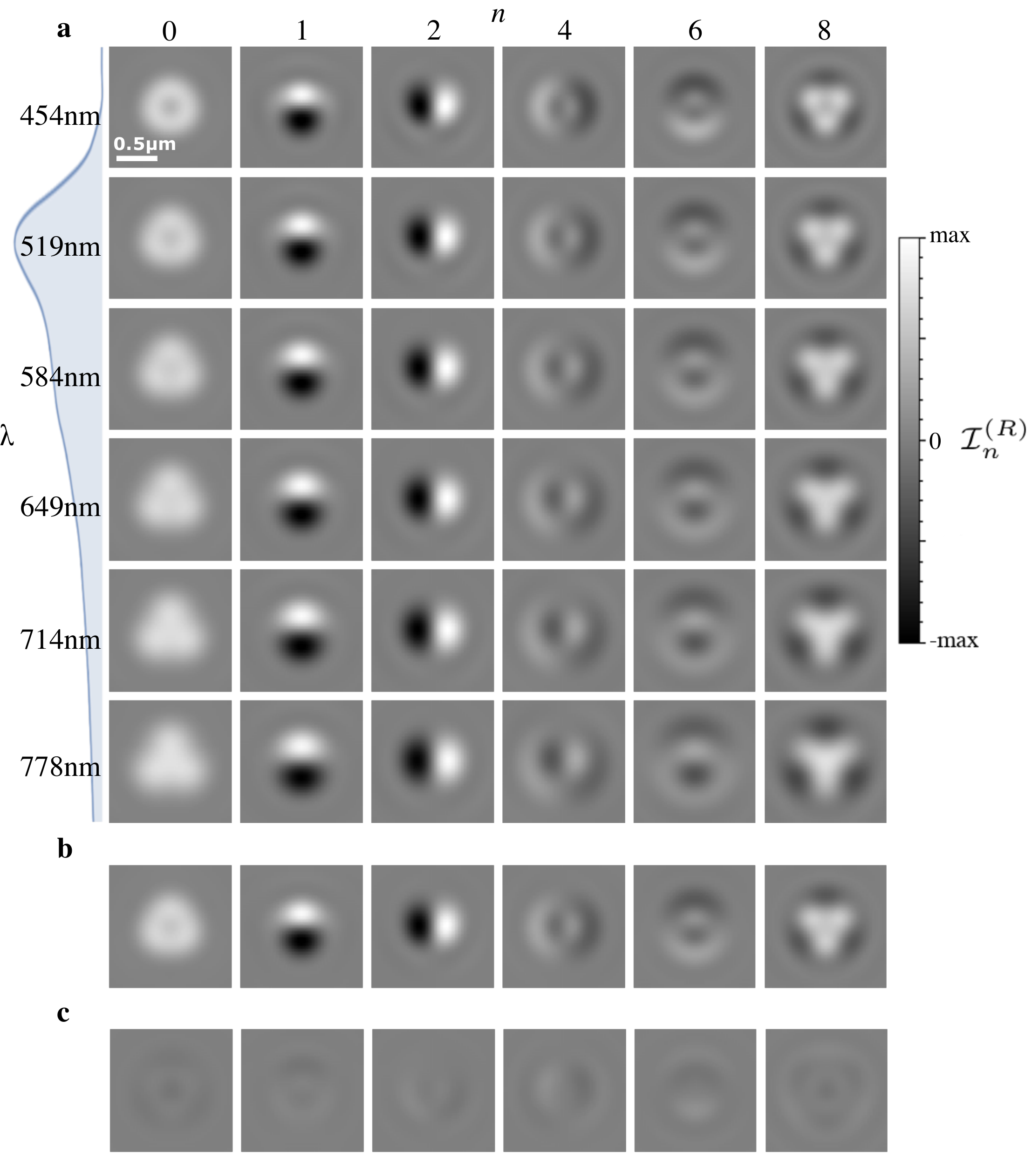}
\caption*{Supplementary Figure 2: Chromatic dependence of the PSF components. (a) Theoretically calculated PSF components ${\cal I}_{n}^{({\mathcal{R}})}$ at $z=0$ for a broad range of wavelengths, where $c=\pi$ at the nominal wavelength of 519nm. Each row is normalized separately. The fluorescence spectrum is shown on the left. (b) Superposition of these PSF components weighted by the whole fluorescence spectrum.  (c) Difference between the polychromatic PSFs in (b) and those for the nominal (peak) wavelength shown in the second row of (a). }
\label{Achroma}
\end{figure}

As explained in the main text, it is assumed that the fluorescent molecules have no chirality, so that ${\cal I}_3^{(p)}$, ${\cal I}_5^{(p)}$ and ${\cal I}_7^{(p)}$ are not required in the retrieval of the localization, orientation and wobbling of the molecules. Note, however, that as mentioned in the main text it is convenient for the alignment of the system to use measurements of fluorescent beads where a circular polarizer is inserted before the SEO. In this case, the PSFs are dominated by ${\cal I}_0^{(p)}$ and ${\cal I}_3^{(p)}$, which are approximately rotationally symmetric and whose combination produces PSFs that are a donut shape and a spot. These shapes vary slowly with defocus around the nominal plane. The alignment of the system and the calibration of residual birefringence prior to the SEO are then adjusted to maximize the rotational symmetry of the measured PSFs.

\subsection{Axial dependence of the PSF.}\label{SectionAxial}
We now study the dependence of the $z$ coordinate of the emitter on the PSFs and the information they carry. First, notice that this $z$ dependence justifies the choice of separating the image in terms of circular polarization components. For example, if no QWP were used, the Wollaston prism would separate two orthogonal linear polarizations. In this case, however, the PSF components ${\cal I}_3^{(p)}$, ${\cal I}_4^{(p)}$ and ${\cal I}_6^{(p)}$ would vanish for $z=0$, and it would not be possible to determine the corresponding generalized Stokes parameters. On the other hand, decomposing the images in terms of circular polarizations leads to PSFs that depend strongly on all the relevant parameters, as we now see. For right ($\mathcal{R}$) and left ($\mathcal{L}$) circular polarizations, the functions in Eq.~(\ref{G1}) become
\bse
\bea
G_{x}^{({\mathcal{R}/\mathcal{L}})}(\brho)&=&\frac1{\sqrt{2}}{\cal F}(\exp(-\ui kz\gamma)\{{\cal C}[g_0-g_2\exp(\mp2\ui\varphi)]\nonumber\\
&+&\ui{\cal S}\exp(\pm\ui\varphi)[g_0-g_2\exp(\pm2\ui\varphi)]\}),\\
G_{y}^{({\mathcal{R}/\mathcal{L}})}(\brho)&=&\mp\frac1{\sqrt{2}}{\cal F}(\exp(-\ui kz\gamma)\{\ui{\cal C}[g_0+g_2\exp(\mp2\ui\varphi)]\nonumber\\
&+&{\cal S}\exp(\pm\ui\varphi)[g_0+g_2\exp(\pm2\ui\varphi)]\}),\\
G_{z}^{({\mathcal{R}/\mathcal{L}})}(\brho)&=&\frac1{\sqrt{2}}{\cal F}\{\exp(-\ui kz\gamma)g_1\nonumber\\
&\times&[{\cal C}\exp(\mp\ui\varphi)+\ui{\cal S}\exp(\pm2\ui\varphi)]\}\eea
\ese
where ${\cal C}=\cos(cu/2)$ and ${\cal S}=\sin(cu/2)$. 
By inserting these expressions into Eqs.~(\ref{PSFcomponents}), one can find the PSF components for any axial displacement $z$. 

While the best retrieval results might be achieved by using the rigorous dependence in $z$ of the model, approximations can be made to find a simplified parametrization in $z$ of the form
\be
{\cal I}_n^{(p)}(\brho,z)\approx\sum_{m=0}^{M}h_{n,m}(z)\,{\cal I}_{n,m}^{(p)}(\brho).
\ee
where the components ${\cal I}_{n,m}^{(p)}$ and the functions $h_{n,m}(z)$ are found by fitting over calculations at different $z$ or from calibration experimental data. For example, for a given $n$, a singular value decomposition can be used to fit ${\cal I}_n^{(p)}$ sampled over several values of $z$ and the leading $M$ terms in this decomposition can be used to approximate the expression. The resulting eigenvectors over the samples in $z$ can be used to find fits for the functions $h_{n,m}(z)$. For small displacements ($|z|\widetilde{<}\lambda/{\rm n}\cos\theta_0$) it can be sufficient to use monomials $h_{n,m}(z)=z^m$, the constant and the linear functions being enough to capture the main features of the transformation. We used this approach in the proof-of-principle experiments with fluorescent beads simulating molecules with different orientations, where the PSF model was constructed from experimental measurements. A quadratic term was included in order to help avoid the resulting approximate PSFs from containing negative values. 

\section*{Supplementary Note 2: Orientation and wobbling information}

\subsection{Correlation matrix.}
As mentioned earlier, the generalized Stokes parameters can be measured experimentally by finding the coefficients of the PSF basis in order to match as well as possible the measured PSF at the two detector regions. 
From these parameters, we can extract information about the dipole's orientation and wobbling, both in the $xy$-plane and out of it. Consider first the case of a dipole whose orientation wobbles around the $z$ axis with main directions aligned with the $x$ and $y$ axes, so that its direction cosines in the $x$ and $y$ directions have standard deviations $\Delta_1$ and $\Delta_2$, respectively. These standard deviations characterize the half-angles of the elliptical cone of directions within which the molecule wobbles. The components of the correlation matrix are then
\be
\mathbf{\Gamma}_0=|E_0|^2\left(\begin{array}{ccc}\Delta_1^2&0&0\\ 0&\Delta_2^2&0\\ 0&0&1-\Delta_1^2-\Delta_2^2\end{array}\right).
\ee
For a molecule with arbitrary orientation, the generic correlation matrix corresponds to a 3D rotation of this matrix:
\be
\mathbf{\Gamma}=\mathbf{R}\mathbf{\Gamma}_0\mathbf{R}^{\rm T},
\ee
where $\mathbf{R}$ is a $3\times3$ rotation matrix. Since both $\mathbf{\Gamma}_0$ and $\mathbf{R}$ are real, so is $\mathbf{\Gamma}$, which means that $S_3=S_5=S_7=0$. That is, as mentioned earlier, only six generalized Stokes parameters are relevant to this problem, of which one, $S_0=|E_0|^2/\sqrt{3}$, is independent of orientation and hence serves only as normalization for the remaining five nonzero parameters.  
The parameters $S_3$, $S_5$, and $S_7$ could be useful, e.g., in the measurement of chiral molecules for which $\mathbf{\Gamma}$ can be complex, but this is not the case of the molecules studied here.

The retrieval of the dipole's main orientation angles $\theta$ and $\xi$, of the standard deviations $\Delta_1$ and $\Delta_2$, and their corresponding directions of vibration can be achieved by simply finding the eigenvalues and eigenvectors of the estimation of $\mathbf{\Gamma}$ resulting from the measurements. Let these eigenvalues and eigenvectors be denoted by $\Lambda_i$ and ${\bf v}_i$, respectively, for $i=1,2,3$, and ordered such that $\Lambda_1\ge\Lambda_2\ge\Lambda_3$. Ideally, the molecule's average orientation is given by ${\bf v}_1$, and $\Delta_1^2=\Lambda_2/T$, $\Delta_2^2=\Lambda_3/T$, where $T={\rm Tr}(\mathbf{\Gamma})=\Lambda_1+\Lambda_2+\Lambda_3$. The directions of oscillation associated with these two variances are those of ${\bf v}_2$ and ${\bf v}_3$, respectively, so it is possible in theory to estimate the asymmetry of the wobbling. In practice, however, limitations due to small numbers of photons, additive noise and pixelation introduce errors in these measures. The effect of these sources of error on the eigenvectors and eigenvalues ${\bf v}_i,\Lambda_i$ is larger for larger $i$ (i.e., for smaller $\Lambda_i$), which means that the estimation of the main orientation ${\bf v}_i$ is significantly more robust than that of the wobbling, particularly for small wobbling angles. Ideally,  the polarization matrix should be non-negative definite, i.e., $\Lambda_i\ge0$. However, the sources of error mentioned earlier can make the estimated $\Lambda_3$ and sometimes even $\Lambda_2$ negative. 
To alleviate this problem, we make the assumption that the wobbling of the dipole is isotropic with respect to the average dipole direction (that is $\Delta_1=\Delta_2=\Delta$) so we use the largest (and more numerically robust) of the two estimated eigenvalues, namely $\Delta^2=\Lambda_2/T$. In this case, the directions ${\bf v}_{2,3}$ are no longer necessary, and the only meaningful directional parameters are the polar and azimuthal angles, $\theta$ and $\xi$, characterizing the direction of the dipole, as well as the amount of wobbling, $\Delta$. For ease of interpretation, this last quantity can be transformed into a cone angle $\delta$ or a solid angle $\Omega$, corresponding to the assumption that the dipole wobbles within this cone with equal probability of being in any direction. A straightforward calculation gives
\be
\frac{\Lambda_2}T=\frac38-\frac{[\cos(\delta/2)+1/2]^2}6=\frac38-\frac{[3/2-\Omega/2\pi]^2}6,
\ee
from where we can find
\be
\Omega=3\pi\left(1-\sqrt{1-\frac83\frac{\Lambda_2}T}\right),
\label{coneangle}
\ee
so that $\Lambda_2=0$ corresponds to no wobbling ($\delta,\Omega=0$) while the opposite extreme of the largest possible value $\Lambda_2=T/3$ (assuming $\Lambda_3=\Lambda_2$) gives isotropic wobbling in all directions corresponding to $\delta=180^\circ$ and $\Omega=2\pi$. Note that in theory the largest value $\Lambda_2$ can take is $T/2$, which violates the assumption of isotropic wobbling but is numerically possible even for isotropic wobbling due to noise or to errors in the reference PSF basis. To prevent the unphysical results for $\Omega$ that Eq.~(\ref{coneangle}) would give, we constrain the values of $\Lambda_2/T$ to the interval $[0,1/3]$, so that when the retrieved value is outside this interval we simply use the closest value within it.
Also, we found through simulations that the results can be improved if an extra step is added, consisting of using the retrieved values as starting points for maximizing the correlation of the measured PSFs to the model constrained to isotropic wobbling. The approach of maximization of the correlation was used for retrieving the parameters in the STORM measurements.

\subsection{Degree of polarization and rotational mobility.}

The so-called rotational mobility or rotational constraint, which describes the amount of wobble of a single fluorophore, has been studied both in its 2D\cite{Backer2016,Zhang2019} and 3D\cite{Backer2015,Zhang2019,Zhang2018} forms. Following the notation in those references, we denote this parameter here as  $\gamma_{\mathrm{2D/3D}}$, depending on the dimensionality. The upper bound for this parameter, $\gamma_{\mathrm{D}}=1$, implies that the molecule is completely fixed, or at least that the wobbling is in a time scale much larger than the integration time of the detector. On the contrary, the lower bound $\gamma_{\mathrm{D}}=0$ means that the molecule wobbles freely and isotropically within the integration time of the detector. 
The relation between the measures of polarization $P_{\rm 2D/ 3D}$ 
and the mobility parameter, $\gamma_{\rm 2D/3D}$, is now described. As mentioned earlier, the second-moment (or polarization) matrix 
$\boldsymbol{\Gamma}$ (denoted as $\mathbf{M}$ by other authors) can be written as 
\begin{equation}
\mathbf{\Gamma}=\sum_{j=1}^{3}\Lambda_j\mathbf{v}_j\mathbf{v}_j^\dagger.
\end{equation}

Let us first consider the 2D case in which the molecule is taken to wobble only within the $xy$-plane. We can then use the the submatrix
\begin{equation}
\mathbf{\Gamma}_{xy}=\sum_{j=1}^{2}{\tilde{\Lambda}}_j\mathbf{\tilde{v}}_j\mathbf{\tilde{v}}_j^\dagger,
\end{equation}
where $\tilde{\Lambda}_j$ are the eigenvalues of the sub-matrix, $\mathbf{\tilde{v}}_j$ their corresponding eigenvectors, and we assume $\tilde{\Lambda}_1\geq\tilde{\Lambda_2}$. Notice that in general $\tilde{\Lambda}_j\neq\Lambda_j$. We can express our submatrix as\cite{Backer2016,Zhang2019}
\begin{equation}
\mathbf{\Gamma}_{xy}=(\tilde{\Lambda}_1+\tilde{\Lambda}_2)\left\{ \gamma_{\mathrm{2D}}(\mathbf{v}_1\mathbf{v}_1^\dagger)+\frac{(1-\gamma_{\mathrm{2D}})}{2}\mathbf{I}\right\},\label{eq:decomposition2D}
\end{equation}
where the rotational mobility parameter in 2D is defined as
\begin{equation}
 \gamma_{\mathrm{2D}}=\frac{\tilde{\Lambda}_1-\tilde{\Lambda}_2}{\tilde{\Lambda}_1+\tilde{\Lambda}_2}
\end{equation}
which corresponds exactly to the definition of degree of polarization for paraxial fields\cite{Brosseau}, $P_{\mathrm{2D}}$. Further, if the dipole wobbles uniformly within an angle\cite{Backer2016} $\delta_{\mathrm{2D}}$, the mobility parameter ends up being
\begin{equation}
\gamma_{\mathrm{2D}}=\frac{\sin\delta_{\mathrm{2D}}}{\delta_{\mathrm{2D}}}.
\end{equation}

The situation changes when considering the 3D problem. For this scenario, a decomposition for $\mathbf{\Gamma}$ analogous to that in Eq. (\ref{eq:decomposition2D}) was proposed\cite{Zhang2019}:
\begin{equation}\label{eq:decomposition3D}
\mathbf{\Gamma}=(\Lambda_1+\Lambda_2+\Lambda_3)\left\{ \gamma_{\mathrm{3D}}(\mathbf{v}_1\mathbf{v}_1^\dagger)+\frac{(1-\gamma_{\mathrm{3D}})}{3}\mathbf{I}+\frac{\Lambda_2-\Lambda_3}{2(\Lambda_1+\Lambda_2+\Lambda_3)}(\mathbf{v}_2\mathbf{v}_2^\dagger-\mathbf{v}_3\mathbf{v}_3^\dagger)\right\},
\end{equation}
where the 3D rotational mobility parameter is given by
\begin{equation}\label{eq:gamma3D}
\gamma_{\mathrm{3D}}=\frac{2\Lambda_1-\Lambda_2-\Lambda_3}{2(\Lambda_1+\Lambda_2+\Lambda_3)}.
\end{equation}
The first term in Eq. (\ref{eq:decomposition3D}) corresponds to the contribution of a completely fixed dipole (or main direction of orientation if there is wobble). The second term, which is proportional to the identity matrix, corresponds to the amount of isotropic wobbling. 
The third term characterizes the possible anisotropy in the rotational mobility, meaning that the dipole would wobble inside a cone with elliptical profile rather than circular.

Notice that this rotational mobility does not correspond in general to the degree of polarization\cite{Sampson,Barakat,Tero,Alonso:2020geometric} used in the main manuscript, which is defined as
\begin{equation}\label{eq:P3D}
P_{\mathrm{3D}}=\left[\frac{3\,{\rm tr}\boldsymbol{\Gamma}^2}{2\,({\rm tr}\boldsymbol{\Gamma})^2}. -\frac12\right]^{1/2}=\frac{\left(\Lambda_1^2+\Lambda_2^2+\Lambda_3^2-\Lambda_1\Lambda_2-\Lambda_1\Lambda_3-\Lambda_2\Lambda_3\right)^{1/2}}{\Lambda_1+\Lambda_2+\Lambda_3}.
\end{equation}
Nevertheless, if we assume that the dipole wobbles isotropically (a common assumption in the literature\cite{Zhang2019,Backer2016,Backer2015}) within a circular cone subtending an angle $\delta_{\mathrm{3D}}$, or solid angle $\Omega_{\mathrm{3D}}$, the two lowest eigenvalues coincide and Eqs.~(\ref{eq:gamma3D}) and (\ref{eq:P3D}) agree:
\begin{align*}
\Lambda_2=\Lambda_3\implies \gamma_{\mathrm{3D}}=P_{\mathrm{3D}}.
\end{align*}
By using Eqs. (\ref{eq:P3D}) and (\ref{eq:gamma3D}) we can find the general relation between these two parameters:
\begin{equation}
P_{\mathrm{3D}}=\gamma_{\mathrm{3D}}\sqrt{1+3\left(\frac{\Lambda_2-\Lambda_3}{2\Lambda_1-\Lambda_2-\Lambda_3}\right)^2},
\end{equation}
which indicates that $P_{\mathrm{3D}}\ge\gamma_{\mathrm{3D}}$, the equality holding only for wobbling that is isotropic (in the second-order sense) around a main direction.

\section*{Supplementary Note 3: Theoretical study of precision and accuracy} 
\subsection{Cram\'er-Rao bound analysis.}
\label{SectionCRB}
We now study theoretically the precision and accuracy of CHIDO. To do this, we first study the Cram\'er-Rao (CR) lower bounds for the estimated parameters. Let us start by considering only the directional parameters. 
Let ${\cal I}$ denote the intensity distribution over the two detectors, which is given by 
\begin{align}
{\cal I}=\sum_{n=0}^8S_n{\cal I}_n,\label{intensityApp}
\end{align}
where, as mentioned earlier, $S_3=S_5=S_7=0$. 
The sum over all pixels on the two channels gives
\begin{align}
\langle{\cal I}\rangle=\sum_{n=0}^8S_n\langle{\cal I}_n\rangle\approx S_0\langle{\cal I}_0\rangle,
\end{align}
where in this section angular brackets denote a sum over all pixels. In the last step 
we used the fact that $\langle{\cal I}_{1,2}\rangle=0$ due to the anti-parity of these PSF elements over each channel (${\cal R}$ and ${\cal L}$), $\langle{\cal I}_{4,6}\rangle=0$ due to the change of sign for the corresponding PSF elements between the channels ${\cal R}$ and ${\cal L}$, and for appropriate choices of $c$ the positive and negative contributions in ${\cal I}_8$ balance sufficiently well that $|\langle{\cal I}_8\rangle|$ is sufficiently smaller than $\langle{\cal I}_0\rangle$. For example, for $c=1.2\pi$, $|\langle{\cal I}_8\rangle|$ is about a fifth of $\langle{\cal I}_0\rangle$, which while not orders of magnitude smaller,  is sufficiently small to make the simple estimates that result from neglecting it meaningful.  

Let us consider first the case in which there is no background intensity. The probability density distribution ${\cal P}$ as a function of the normalized Stokes parameters $s_n=S_n/S_0$ is 
\begin{align}
{\cal P}\approx\frac{{\cal I}_0+\sum_{n=1}^8s_n{\cal I}_n}{\langle{\cal I}_0\rangle},\label{probApp}
\end{align}
Before finding the Fisher information with respect to the angular parameters $\xi,\theta,\Omega$, it is useful to estimate the Fisher information matrix with respect to the normalized generalized Stokes parameters $s_n$. Given the linear dependence of ${\cal I}$ on these parameters, we find that the elements of this matrix are
\begin{align}
{\cal J}_{s_ns_{n'}}^{\rm Stokes}&={\cal N}\left\langle\frac{\partial_{s_n}{\cal P}\,\partial_{s_{n'}}{\cal P}}{{\cal P}}\right\rangle\approx\frac{{\cal N}}{\langle{\cal I}_0\rangle}\left\langle\frac{{\cal I}_n{\cal I}_{n'}}{{\cal I}_0+\sum_{n=1}^8s_n{\cal I}_n}\right\rangle,
\end{align}
for $n,n'=1,2,...,8$, and where ${\cal N}$ is the number of photons in the measured PSF. 
It turns out that a good order-of-magnitude estimate can be obtained by ignoring the part in the denominator that depends on $s_n$, leading to
\begin{align}
{\cal J}_{s_ns_{n'}}^{\rm Stokes}&\approx\frac{{\cal N}}{\langle{\cal I}_0\rangle}\left\langle\frac{{\cal I}_n{\cal I}_{n'}}{{\cal I}_0}\right\rangle,\label{FApp}
\end{align}
From this approximation, we can appreciate the usefulness of the PSF components ${\cal I}_n$ being approximately orthogonal (for appropriate choices of $c$): the Fisher information matrix is then approximately diagonal, namely
\begin{align}
{\cal J}^{\rm Stokes}&\approx{\cal N}{\rm Diag}(a_1,a_2,a_3,a_4,a_5,a_6,a_7,a_8).
\end{align}
For the values of $c$ used here, the only nondiagonal elements of the Fisher information matrix that are not orders of magnitude smaller than the diagonal ones are ${\cal J}^{\rm Stokes}_{15}$ and ${\cal J}^{\rm Stokes}_{27}$, which are not relevant here since $s_5=s_7=0$. For $c=1.2\pi$, the relevant diagonal elements are approximately $a_1\approx a_2\approx0.63$, $a_4\approx a_6\approx0.41$, and $a_8\approx0.47$. 

Note that $s_n$ can be parametrized in terms of the directional parameters $\xi,\theta,\Omega$ as
\begin{subequations}
\begin{align}
s_1&=\frac{\sqrt{3}}2 P_{\rm 3D} \sin^2\theta\cos2\xi,\\
s_2&=\frac{\sqrt{3}}2 P_{\rm 3D} \sin^2\theta \sin2\xi,\\
s_4&=\frac{\sqrt{3}}2 P_{\rm 3D} \sin2\theta \cos\xi,\\
s_6&=\frac{\sqrt{3}}2 P_{\rm 3D} \sin2\theta\sin\xi,\\
s_8&= -P_{\rm 3D} \frac{1 + 3 \cos2\theta}4.
\end{align}
\end{subequations}
We can then calculate the Fisher information matrix for the directional variables as
\begin{align}
{\cal J}_{\alpha_i\alpha_{i'}}^{\rm angular}&=\sum_{n,n'}{\cal J}^{\rm Stokes}_{s_ns_{n'}}(\partial_{\alpha_i} s_{n})\,(\partial_{\alpha_{i'}} s_{n'}),
\end{align}
where $\alpha_{1,2,3}=\xi,\theta,\Omega$. By using the diagonal approximation for ${\cal J}^{\rm Stokes}$, we can find very simple approximate expressions for the angular standard deviations:
\begin{subequations}
\begin{align}
\sigma_\xi&\approx\frac{(4\pi)^2}{(8\pi^2-6\pi\Omega+\Omega^2)\sin\theta\sqrt{6{\cal N}(u_+-u_-\cos2\theta)}},\\
\sigma_\theta&\approx\frac{2(4\pi)^2}{(8\pi^2-6\pi\Omega+\Omega^2)\sqrt{6{\cal N}(v_+-v_-\cos4\theta)}},\\
\sigma_\Omega&\approx\frac{2(4\pi)^2}{(3\pi-\Omega)\sqrt{2{\cal N}(w-12u_-\cos2\theta+3v_-\cos4\theta)}},
\end{align}
\end{subequations}
with $u_\pm=a_1\pm a_4$, $v_\pm=a_1\pm4a_4+3a_8$, and $w=9a_1+12a_4+11a_8$. Notice that these estimates are independent of $\xi$ and, if all $a_n$ were equal, the dependence in $\theta$ of these estimates would disappear except for the inverse dependence on $\sin\theta$ for $\sigma_\xi$. Even when these coefficients are not exactly equal, the terms involving $u_-,v_-$ are considerably smaller and can be neglected, leading to even simpler estimates:
\begin{subequations}
\begin{align}
\sigma_\xi&\approx\frac{(4\pi)^2}{(8\pi^2-6\pi\Omega+\Omega^2)\sin\theta\sqrt{6{\cal N}}},\\
\sigma_\theta&\approx\frac{(4\pi)^2}{(8\pi^2-6\pi\Omega+\Omega^2)\sqrt{6{\cal N}}},\\
\sigma_\Omega&\approx\frac{(4\pi)^2}{2.8(3\pi-\Omega)\sqrt{\cal N}}.
\end{align}
\end{subequations}
Note that, in arriving at these simple approximations, we rounded up numerical factors coming from $u_+$ and $v_+$, so that $\sigma_\xi\sin\theta\approx\sigma_\theta$. If the numerical quantities are calculated more precisely one finds that $\sigma_\theta$ is slightly larger than $\sigma_\xi\sin\theta$, and this is reflected in the rigorously computed CR lower bounds presented in the main manuscript. These expressions become even simpler when expressed in terms of $P_{\rm 3D}$:
\begin{subequations}
\begin{align}
\sigma_\xi&\approx\frac{2}{P_{\rm 3D}\sin\theta\sqrt{6{\cal N}}},\\
\sigma_\theta&\approx\frac{2}{P_{\rm 3D}\sqrt{6{\cal N}}},\\
\sigma_{P_{\rm 3D}}&\approx\frac{1.43}{\sqrt{\cal N}}.
\end{align}
\label{eqs:CRdir}
\end{subequations}
The form of these equations suggests an interpretation in terms of a spherical coordinate system, where $\xi$ and $\theta$ are the azimuthal and polar coordinates and $P_{\rm 3D}$ is the radial one. If $\sigma_\xi$, $\sigma_\theta$ and $\sigma_{p_{\rm 3D}}$ are the widths (assumed as small) that determine the precision in each of the three coordinates, the element of volume would be $\sigma_{\rm Dir}=P_{\rm 3D}^2\sin\theta\,\sigma_{P_{\rm 3D}}\,\sigma_\theta\,\sigma_\xi$, which can be found to simplify to
\begin{align}
\sigma_{\rm Dir}&\approx{\cal N}^{-3/2}.
\end{align}

In real measurements, in addition to the measured PSFs there is typically a fairly uniform background. This background also contributes to the noise (assumed here to be Poissonian), and hence must be considered in the calculations for the CR lower bounds. To account for its effect, we add this background to the intensity distribution in Eq.~(\ref{intensityApp}), so that it enters both in the numerator and denominator of the probability density in Eq.~(\ref{probApp}). This modification causes a few changes in the estimate of the Fisher information matrix in Eq.~(\ref{FApp}): first, the number of photons used must be that of the signal plus background, but some of the prefactors that are extracted from the sum over pixels due to the change of normalization of ${\cal P}$ amount to the fraction of all photons that are due only to the signal. The combination of these two changes then can be combined into a new ${\cal N}$ that can interpreted as the number of signal photons. The second effect is that the background must be added to ${\cal I}_0$ in the denominator inside the large angular brackets of the last factor in the right-hand side of Eq.~(\ref{FApp}). Since the average of ${\cal I}_0$ over the region occupied by the PSFs is approximately half its peak value, and since the SBR is defined as the ratio of the peak of the PSF to the background, then this change in normalization can be approximated by a factor of $(1+2\,{\rm SBR}^{-1})^{-1}$. The CR lower bounds then must be multiplied by a factor of $\sqrt{1+2\,{\rm SBR}^{-1}}$. This approximation is validated by rigorous calculations of the six CR lower bounds for both the directional and spatial parameters, as shown in Supplementary Fig.~3(a).
\begin{figure}[htbp]
\centering
\includegraphics[width=0.85\linewidth]{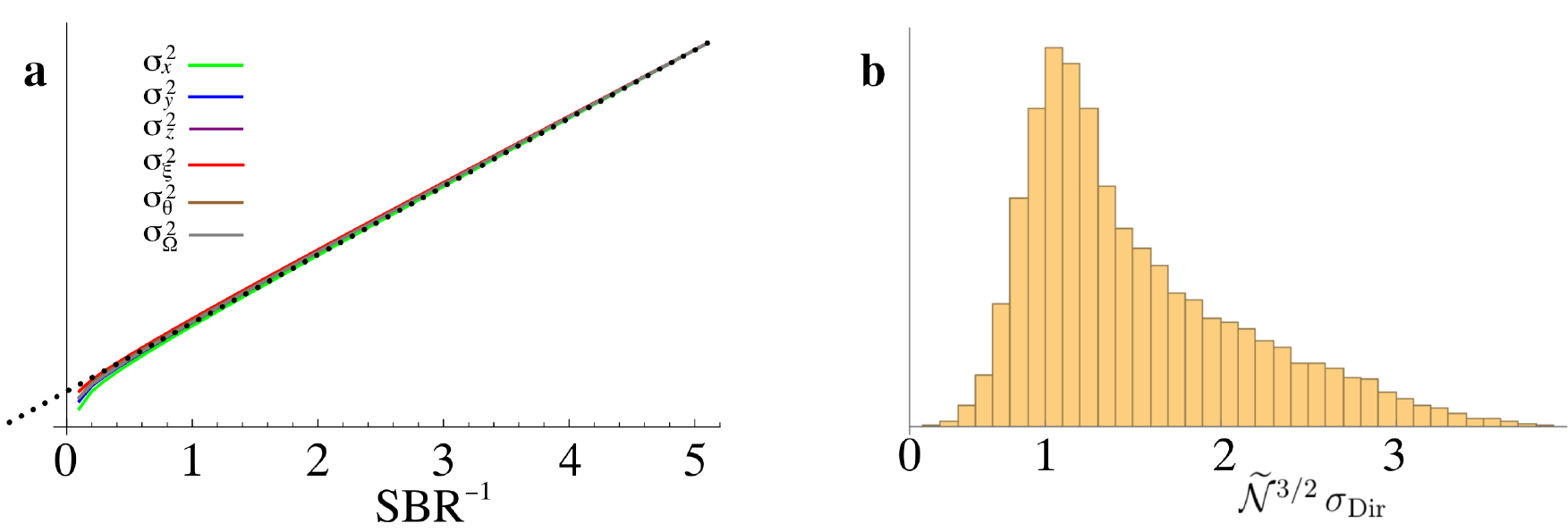}
\caption*{Supplementary Figure 3: Dependence of Cram\`er Rao lower bounds on signal-to-background ratio, and global measure of directional/wobbling precision. (a) Squares of the CR lower bounds for the position and direction parameters, as functions of ${\rm SBR}^{-1}$, for a non-wobbling fluorophore with $\xi=0$, $\theta=\pi/2$, and $x=y=z=0$. All plots were rescaled to show their proportionality to $1+2\,{\rm SBR}^{-1}$ (dotted line). (b) Histogram of $\widetilde{\cal N}^{3/2}\,\sigma_{\rm Dir}$ calculated rigorously for 10000 randomly generated cases with ${\rm SBR}^{-1}\in[0,3]$ and statistically uniform sampling of the directional parameter space $(P_{\rm 3D},\xi,\theta)$.}
\label{SBRscale}
\end{figure}
The expression for the global directional lower bound $\sigma_{\rm Dir}$ then acquires a factor of $(1+2\,{\rm SBR})^{3/2}$, as described in the main body. Supplementary Fig.~3(b) shows a histogram of $\widetilde{\cal N}^{3/2}\,\sigma_{\rm Dir}$ (where $\widetilde{\cal N}={\cal N}/(1+2 {\rm SBR}^{-1})$) calculated rigorously for 10000 randomly generated cases with ${\rm SBR}^{-1}\in[0,3]$, and for statistically uniform sampling of the directional parameter space $(P_{\rm 3D},\xi,\theta)$. We can see that this histogram is indeed peaked near unity.

Let us now consider the Fisher information matrix for both the spatial and directional parameters, ${\cal J}^{\rm all}_{\alpha_i\alpha_{i'}}$, for $\alpha_{1,2,3,4,5,6}=x,y,z,\xi,\theta,\Omega$. The coupling between the parameters can be characterized by a normalized version of this matrix, given by
\be
C^{\rm all}_{\alpha_i\alpha_{i'}}=\frac{{\cal J}^{\rm all}_{\alpha_i\alpha_{i'}}}{\sqrt{{\cal J}^{\rm all}_{\alpha_i\alpha_{i}}{\cal J}^{\rm all}_{\alpha_{i'}\alpha_{i'}}}}.
\label{correq}
\ee
The closer this matrix is to the identity, the lower the level of coupling between the parameters. Supplementary Fig.~4 shows both the average and standard deviation (element by element) of this matrix, averaged over all possible directions and over $z\in$ [$-200$ nm, $200$ nm], for different values of $P_{\rm 3D}$, and both without background and with a SBR of $1/3$. The coupling has negligible systematic bias, so average tends to a diagonal matrix. The standard deviations of the non-diagonal elements do grow when there is more wobble and/or background, but remain, for the chosen parameters, significantly below unity. 

\begin{figure}[htbp]
\centering
\includegraphics[scale=0.68]{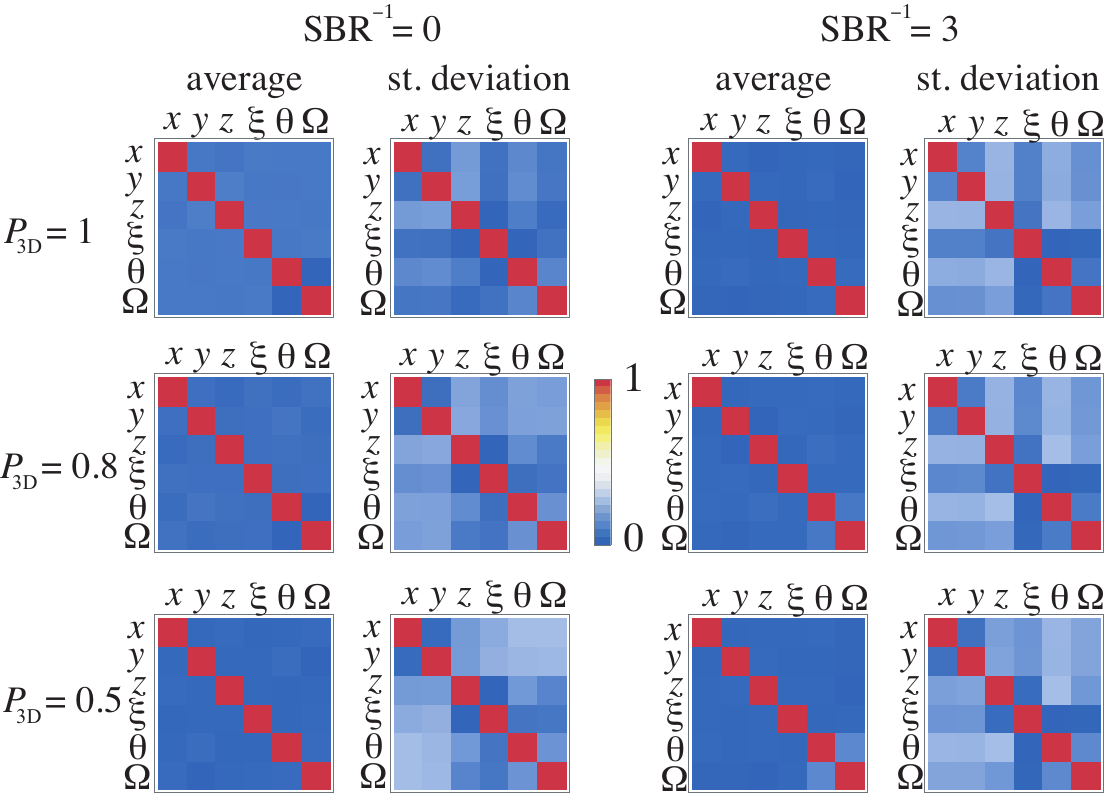}
\caption*{Supplementary Figure 4: Level of coupling in the estimation of different parameters. Averages and standard deviations of the elements of the correlation in Eq.~(\ref{correq}) in the absence of background and for a SBR of $1/3$, for $P_{\rm 3D}=1,0.8,0.5$ (namely $\Omega=0,0.28\pi,0.76\pi$).}
\label{figcorr}
\end{figure}

Finally to show the robustness of CHIDO to aberrations, we repeat the CRB simulations in Fig.~4 of the main manuscript for the case in which the system has one wave of spherical aberrations. The results are shown in Supplementary Fig.~5.
\begin{figure}[htbp]
\centering
\includegraphics[width=0.7\linewidth]{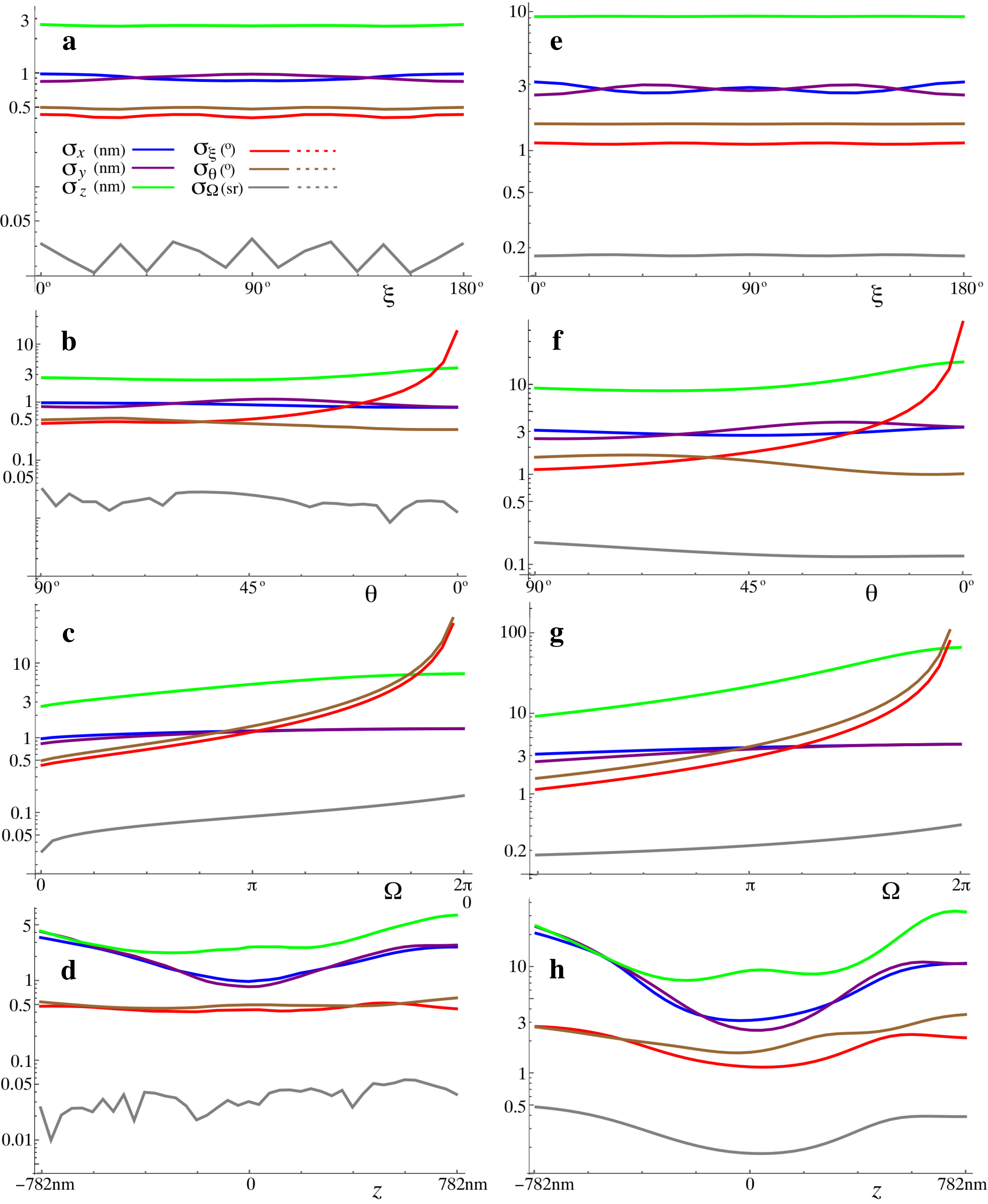}
\caption*{Supplementary Figure 5: Cram\'er-Rao lower bounds for the six measured parameters when the system presents one wave of spherical aberration. These plots assume $10000$ signal photons over the two channels, for (a-d) no background photons, and (e-h) a SBR of $1/3$. The parameters are: (a,e) $x=y=z=0$, $\theta=90^\circ$, $\Omega=0$ and varying $\xi$; (b,f) $x=y=z=0$, $\xi=0^\circ$, $\Omega=0$ and varying $\theta$; (c,g) $x=y=z=0$, $\xi=0^\circ$, $\theta=90^\circ$, and varying $\Omega$; and (d,g) $x=y=0$, $\xi=0^\circ$, $\theta=90^\circ$, $\Omega=0$ and varying $z$. The units for each curve are indicated in the legend in (a). The origin in $z$ was shifted by 300 nm, to the plane where the PSFs are the most localized. }
\label{spherical}
\end{figure}

\subsection{Monte Carlo simulations and assessment of precision and accuracy.}

In order to assess the bias (accuracy) and standard deviations (precision) in the estimation of the parameters intrinsic to the method, we simulated numerically PSF pairs with $10000$ photons for a range of different parameter values. Given the Poissonian nature of the noise, the optimal retrieval method would be the maximization of the likelihood. However, the method used here, which was easier to implement, was the minimization of the RMS error between the model and the simulated noisy PSFs or, equivalently, the maximization of the normalized correlation between them. 

To compare the results with the CR bounds, we considered the same scenarios as in Fig.~4 of the main manuscript, each with 5000 realizations.  Supplementary Fig.~6(a-d) shows the standard deviations of the estimations with no background, while (e-h) show the corresponding results for a SBR of $1/3$. Even though the estimation method is not ideal for Poisson noise, the precision obtained agrees well with that predicted from the CR bound analysis, the measured standard deviations being in most cases within a factor of about 3 from these bounds. Note that, for the wobble angle $\Omega$ the standard deviations from the parameter retrieval are sometimes smaller that the CR bound. This is because the range for this parameter is finite (and non-periodic) and we are considering cases at an edge of this range. The results are then squeezed against the edge of the allowed interval, giving a distribution that is narrower than what is predicted by the CR analysis. As discussed in what follows, proximity to the edge also causes a bias away from it, comparable in size to the standard deviation.

\begin{figure}[htbp]
\centering
\includegraphics[scale=0.55]{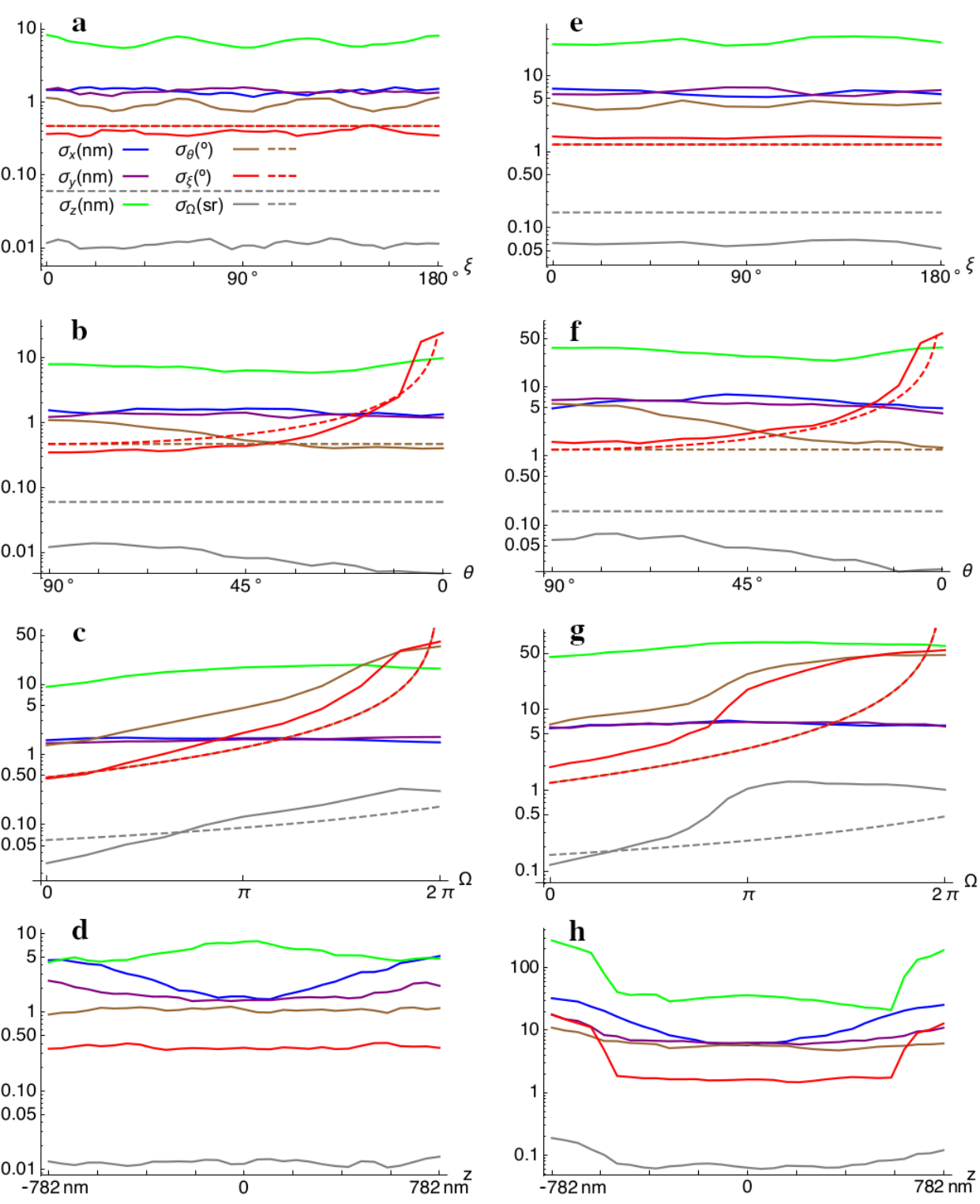}
\caption*{Supplementary Figure 6: Standard deviations for the retrieval of the six measured parameters of numerically-simulated noisy PSFs. These simulations assume $10000$ signal photons over both channels, for (a-d) no background photons and (e-h) a SBR of $1/3$. For each scenario 5000 random realizations were simulated. The parameters are: (a,e) $x=y=z=0$, $\theta=90^\circ$, $\Omega=0$ and varying $\xi$; (b,f) $x=y=z=0$, $\xi=0^\circ$, $\Omega=0$ and varying $\theta$; (c,g) $x=y=z=0$, $\xi=0^\circ$, $\theta=90^\circ$, and varying $\Omega$; and (d,g) $x=y=0$, $\xi=0^\circ$, $\theta=90^\circ$, $\Omega=0$ and varying $z$. The units for each curve are indicated in the legends of (a,e). The dashed lines indicate the simple estimates given in Eqs.~(\ref{eqs:CRdir}). }
\label{simulations}
\end{figure}

Regarding accuracy, no statistically meaningful biases were found, the average retrieved values being typically of the order of the standard deviation divided by the square root of the number of test cases being averaged. The average deviations from the true value are shown in Supplementary Fig.~7 for the same eight cases as in Supplementary Fig.~6. As mentioned earlier, only for $\Omega$ do we see significant biases, but these are caused by the fact that this parameter is defined within a finite interval: for values of this parameter within a few standard deviations from the edges of the interval, there is a small bias away from the edges, of the order of the standard deviation, since the estimation only allows errors to one side of the edges. 
Note that these figures were generated by using a dense sample of values of the parameter in question, and then averaging cases within small intervals.

\begin{figure}[htbp]
\centering
\includegraphics[scale=0.55]{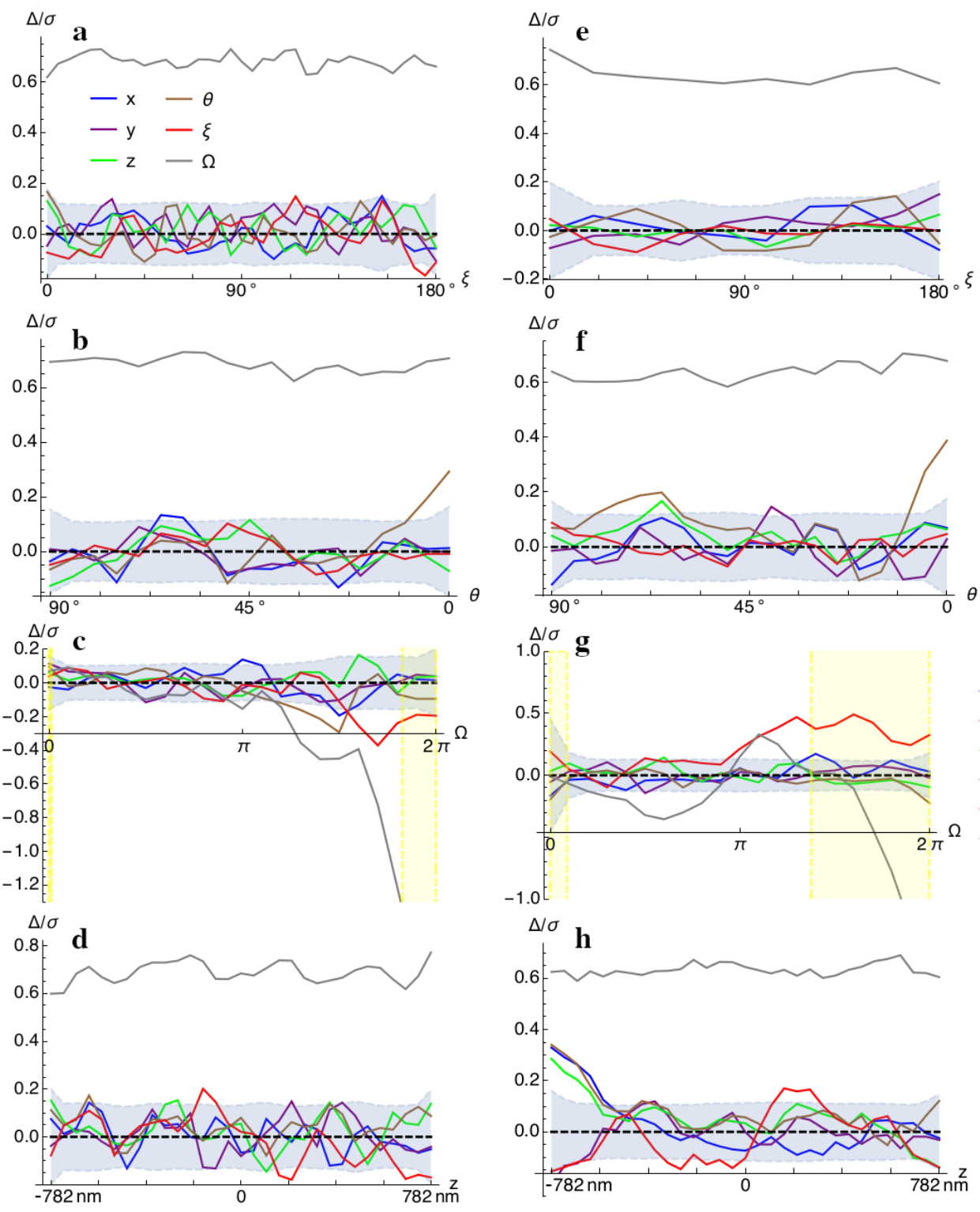}
\caption*{Supplementary Figure 7: Assessment of bias in the estimation of the parameters. Averaged deviation from the true value for each parameter,
normalized by the corresponding standard deviation, for each of the cases presented in Supplementary Fig.~6.
The dashed thick line indicates zero bias. The blue region corresponds to $\pm 2/\sqrt{n_{\mathrm{samples}}}$, and values that do not deviate far beyond this region can be regarded as statistical fluctuations rather than biases. For (c,g) the yellow regions indicate values of the parameter within $ 2\sigma_\Omega$ from the edges of the allowed interval for $\Omega$. }
\label{simulationsbias}
\end{figure}

\section*{Supplementary Note 4:  Retrieval of parameters}\label{SectionRetrieval}
In this supplementary note we describe a fast approach for the retrieval of the parameters when a basis of PSFs associated with the generalized Stokes parameters and the coefficients of an expansion in $z$ is used. If more accurate results are needed, these parameter estimates can be used as starting points in a more rigorous parameter retrieval routine through the maximization of the likelihood function or the minimization of the rms error. 
Supplementary Fig.~8(a) shows a theoretically calculated basis for $p=\mathcal{R}$ for a quadratic ($M=2$) expansion in $z$. The corresponding PSFs for $p=\mathcal{L}$ are the same except for a sign change in some of its components (encloded in red in the figure). Note that the SEO's orientation was chosen to coincide with the orientation used in the bead measurements, for which the estimated basis, shown in Supplementary Fig.~8(b), was constructed according to a method discussed later.

\begin{figure}[htbp]
\centering
\includegraphics[scale=0.5]{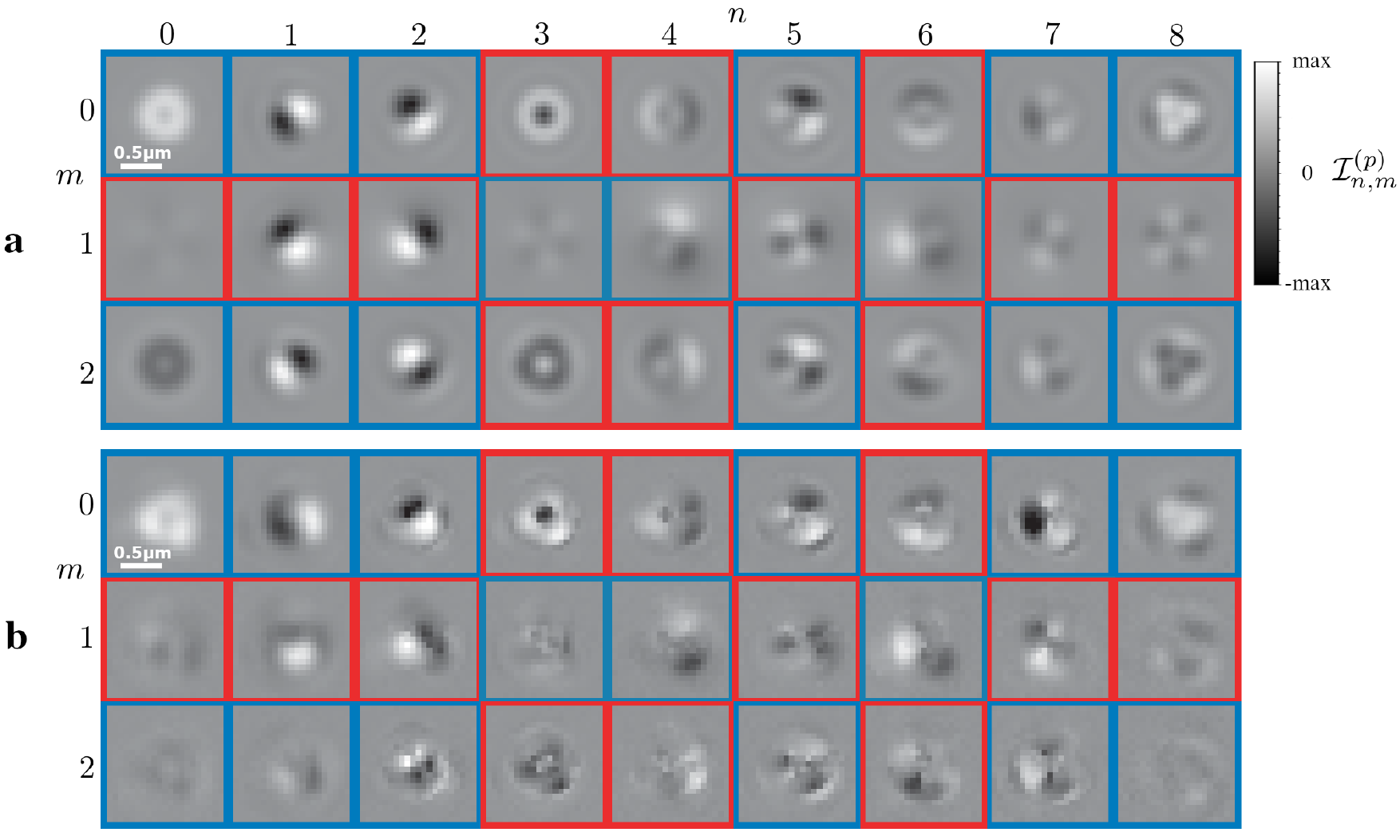}
\caption*{Supplementary Figure 8: Comparison of a theoretical PSF basis and one constructed from a mixture of reference measurements with fluorescent beads and theory. (a) Theoretically calculated PSF components ${\cal I}_{n,m}^{(p)}$, with $m=0$ for PSFs at $z=0$, for the detector capturing the RHC polarization components ($p={\mathcal{R}}$), using $c=\pi$ and assuming $t_{\rm s}=t_{\rm p}=1$. For LHC polarization ($p={\mathcal{L}}$), the PSF components are the same except that those enclosed by red frames change sign, while the ones enclosed by blue frames remain unchanged. 
Each row is normalized separately, as the units are different.   The SEO is rotated by an angle of $-54.4^\circ$, which fits with the experimental measurements. (b) Corresponding PSF components by using a mixture of experimental and theoretical data.} 
\label{PSFs_system}
\end{figure}

\subsection{Transverse localization.}
The determination of transverse position is performed by working in the (discrete) Fourier domain, which also facilitates accounting for the pixelization of the images. The idea is to find the superposition of displaced versions of the PSFs that agrees the most with the measured intensity. This translates into finding the dispacements $x,y$ and the coefficients $S_{n,m}$ for the PSF basis elements ${\cal I}_{n,m}^{(p)}$ that minimize the merit function
\be
\mu=\sum_{p={\rm r,l}}\left\langle\left(\widehat{\cal T}_{-x,-y}I^{(p)}-\sum_{n,m}S_{n,m}{\cal I}_{n,m}^{(p)}\right)^2\right\rangle,
\label{meritfunction}
\ee
where $\langle\cdot\rangle$ denotes sum over all pixels, and $\widehat{\cal T}_{x,y}$ indicates a translation in $x$ and $y$, which can be implemented in Fourier space as a linear phase and is therefore not constrained to integer multiples of the pixel size. Note that, for convenience, we applied the displacement (with opposite signs) to the measured intensity rather than to the PSF basis functions. 

Consider first the solution for the coefficients $S_{n,m}$. These are found by setting to zero the derivative of $\mu$ with respect to each of these coefficients, leading to a set of equations of the form
\be
\sum_{n',m'}S_{n',m'}a_{n',m',n,m}=b_{n,m}(x,y),
\label{system}
\ee
where
\bse
\be
a_{n',m',n,m}=\sum_{p={\rm r,l}}\left\langle{\cal I}_{n',m'}^{(p)}{\cal I}_{n,m}^{(p)}\right\rangle
\label{aPSFcorr}
\ee
can be thought of as the elements a $6(M+1)\times6(M+1)$ matrix ${\bf a}$, where $M$ is the maximum value of $m$, and
\bea
b_{n,m}(x,y)&=&\sum_{p={\rm r,l}}\left\langle{\cal I}_{n,m}^{(p)}\widehat{\cal T}_{-x,-y}I^{(p)}\right\rangle\nonumber\\
&=&\sum_{p={\rm r,l}}{\cal F}^{-1}[({\cal F}I^{(p)})^*({\cal F}{\cal I}_{n,m}^{(p)})](x,y),
\label{corrs}
\eea
\ese
are the correlation of the measured intensity with each of the basis elements. Note that the left-hand side of Eq.~(\ref{system}) can be interpreted as the product of ${\bf a}$ and the vector whose components are the coefficients $S_{n,m}$. 
This equation 
can  be easily solved for these unknown coefficients by finding ${\bf a}^{-1}$ and multiplying it by the vector whose elements are 
$b_{n,m}(x,y)$, namely
\be
S_{n,m}=\sum_{n',m'}b_{n',m'}(x,y)\{{\bf a}^{-1}\}_{n',m',n,m}.
\label{solutions}
\ee
Finally, notice that the substitution of this solution into Eq.~(\ref{meritfunction}) gives
\be
\mu=\sum_{p={\rm r,l}}\langle [I^{(p)}]^2\rangle-f(x,y),
\ee
where the explicitly real and positive function $f$ is defined as
\be
f(x,y)=\sum_{n,m,n',m'}\{{\bf a}^{-1}\}_{n',m',n,m}\,b_{n,m}^*(x,y)\,b_{n',m'}(x,y).
\ee
Therefore, the merit function $\mu$ is minimized by maximizing the function $f(x,y)$ in $x$ and $y$. Note that this function is given in terms of the correlations in Eq.~(\ref{corrs}), which can be calculated through fast Fourier transforms. The location in $x,y$ of the maxima (corresponding to the transverse position of the emitter) can be determined with accuracy well below a pixel by interpolating via zero padding in the Fourier domain and/or by using a polynomial fit using the values of the pixels surrounding the one with the maximum value. 

Finally, notice that the procedure just described allows finding the centroids of multiple emitters in an image (as long as these are well separated), whose positions are given by the local maxima of $f(x,y)$. Once these coordinates are found for each emitter, they can be substituted into Eq.~(\ref{solutions}) to find the coefficients $S_{n,m}$ and from them the position in $z$ and the orientation and wobbling of the emitter (see next Section). For this final retrieval of the parameters, it is a good idea to use only the section of the images that contains the measured PSF in question. Finally, $\mu$ (after appropriate normalization) provides a measure of the quality of the fit, and can therefore be used as a measure of confidence in the results. This measure was used to filter out result where the PSFs overlapped, were too noisy, or were clipped by the edge of the field of view.

\subsection{Generation of a PSF bases for bead measurements.}\label{SectionRefs}
As discussed in the main manuscript, fluorescent beads in combination with either a linear polarizer  or a radial polarization waveplate (S-waveplate) before the SEO were used to simulate molecule orientations within the $xy$-plane and in the $z$ direction, respectively. 
For the first case, the reference PSFs were obtained by choosing a particularly bright and well isolated bead from the second set of measurements. 
Measurements were taken for different orientations of the linear polarizer over a range of $180^\circ$, at steps of $10^\circ$ (a total of 19 measurements), and at five defocus distances with a spacing of 200 nm. The polarizer was placed not far from the pupil plane, where light is collimated so that the small amount of wedge in the polarizer does not cause changes in defocus as it is rotated. 
Instead, this wedge did cause a displacement of the PSFs, which moved along a semicircle as the polarizer was rotated by $180^\circ$. 
Because the initial and final orientations of the polarizer correspond to the same polarization and hence give rise to the same shape of the PSFs, it was easy to determine the length and orientation of the diameter joining the endpoints of this semicircular path by correlating the initial and final PSFs. 
The displacement could then be removed computationally (through multiplication by appropriate linear phases in the Fourier domain, so that displacements by fractions of a pixel were possible), leading to a set of PSFs whose origins are consistent. 
After this recentering, an array of $21\times21$ pixels containing the PSFs was selected for each. The 40 pixels at the edge of these arrays were used to calculate the background level of the measurements, which was then subtracted. 
Also, because the number of photons fluctuated from measurement to measurement, each of the 19 arrays was renormalized to make it consistent with the others. 
From these 19 measurements a fit was performed that predicted the intensity distribution of any orientation in the $xy$-plane, and from it it was possible to calculate the corresponding PSF elements ${\cal I}_1^{(p)}$ and ${\cal I}_2^{(p)}$. This procedure was repeated for all five sets of defocus measurements.

Similarly, a sub-basis was generated that emulates the PSFs of fluorophores normal to the plane, by using a bead chosen from one of the samples (set 3) measured with an S-waveplate at the pupil. Again, an array of $21\times21$ pixels containing the PSFs was selected for each of the five defocus measurements (also spaced by 200 nm), and the background was subtracted by using the values of the edge pixels.

Approximating the dependence in $z$ of the measured PSFs by a simple quadratic polynomial does lead to errors in the estimation. However, these errors are largely systematic and can be corrected by using reference measurements. Supplementary Fig.~9(a) and Supplementary Movie 2  
show the averages and standard deviations of the raw estimates of $z$ for the four sets of beads imaged with an S-waveplate to simulate a molecule normal to the object plane. We can see that the spacing of the estimates is underestimated particularly away from $z=0$. In this case, remapping the results through a simple cubic expression leads to the corrected estimates shown in Fig.~5(d) of the main text, which are all spaced by approximately 200 nm. Something similar happens for the bead measurements in which a linear polarizer is used to mimic molecules at different in-plane angles $\xi$. In this case, the distortion caused by the low degree of the polynomial fit also introduces a small amount of coupling between $z$ and $\xi$, as shown by the raw results in Supplementary Fig.~9(b) as well as Supplementary Movies 3 and 4.
Again, because the error is systematic, it can be largely corrected by applying a simple mapping. In this case the mapping was applied to correct only the measurements of set 2, but as can be seen from Fig.~6(b) of the main text, this correction also fixes significantly the estimates of set 1, except for those for the most negative values of $z$ which fell outside the remapped region. 

\begin{figure}[htbp]
\centering
\includegraphics[scale=0.46]{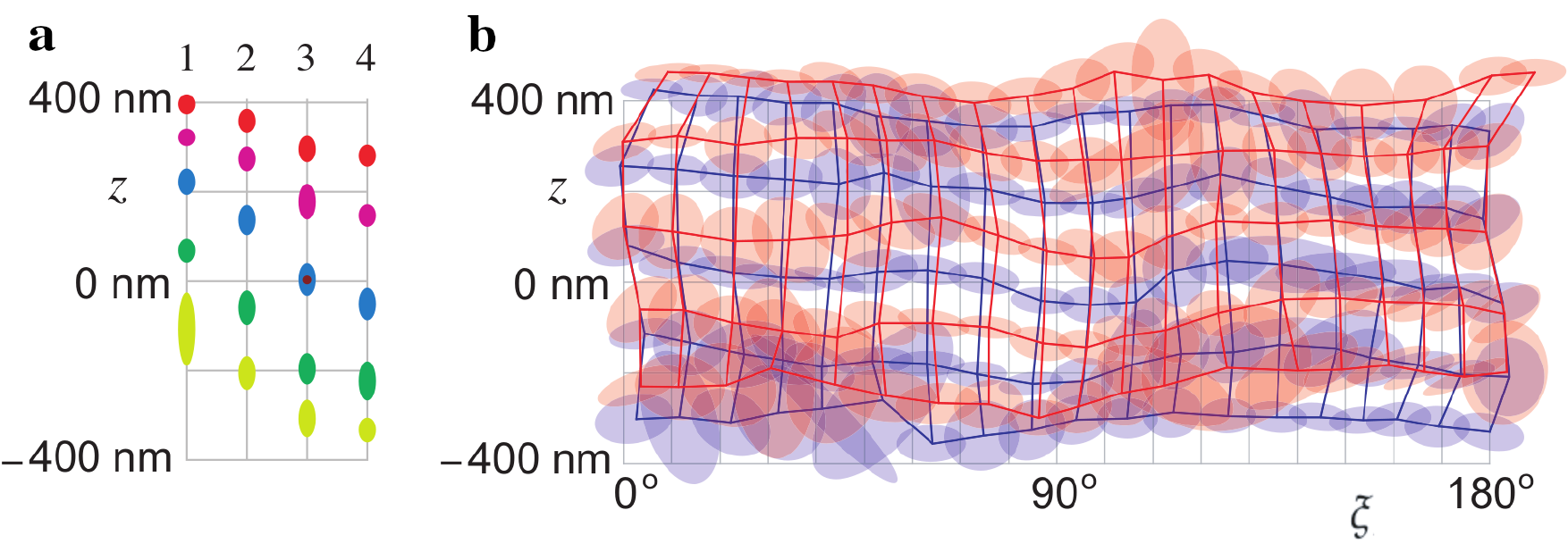}
\caption*{Supplementary Figure 9: Raw estimations of height and in-plane direction for bead measurements. (a) Estimation of $z$ using quadratic approximation (average - center of the ellipse, and standard deviation - height of the ellipse) for the five defocused measurements of the four sets of measurements. All retrieved data are depicted in 
Supplementary Movie 2. 
The  corresponding data fixed by using a cubic correction is shown in Fig.~5(d) of the main text. (b) The intersection points of the blue and red grids indicate the averages of the raw retrieved heights and orientation angles for each measurement, for sets 1 and 2, respectively, and the ellipses centered at each intersection indicate the corresponding standard deviations. 
A shift of $16^\circ$ was applied to the $\xi$-axis so that the retrieved angles fall in the range $[0^\circ,180^\circ]$ for ease of interpretation. The full set of data is shown in Supplementary Movie 3 for set 1 and Supplementary Movie 4 for set 2. 
The corresponding results after the application of a correction to calibrate the results of set 2 (red) is shown in Fig.~6(b) of the main text.}
\label{unmapped}
\end{figure}

\begin{figure}[htbp]
\centering
\includegraphics[scale=.45]{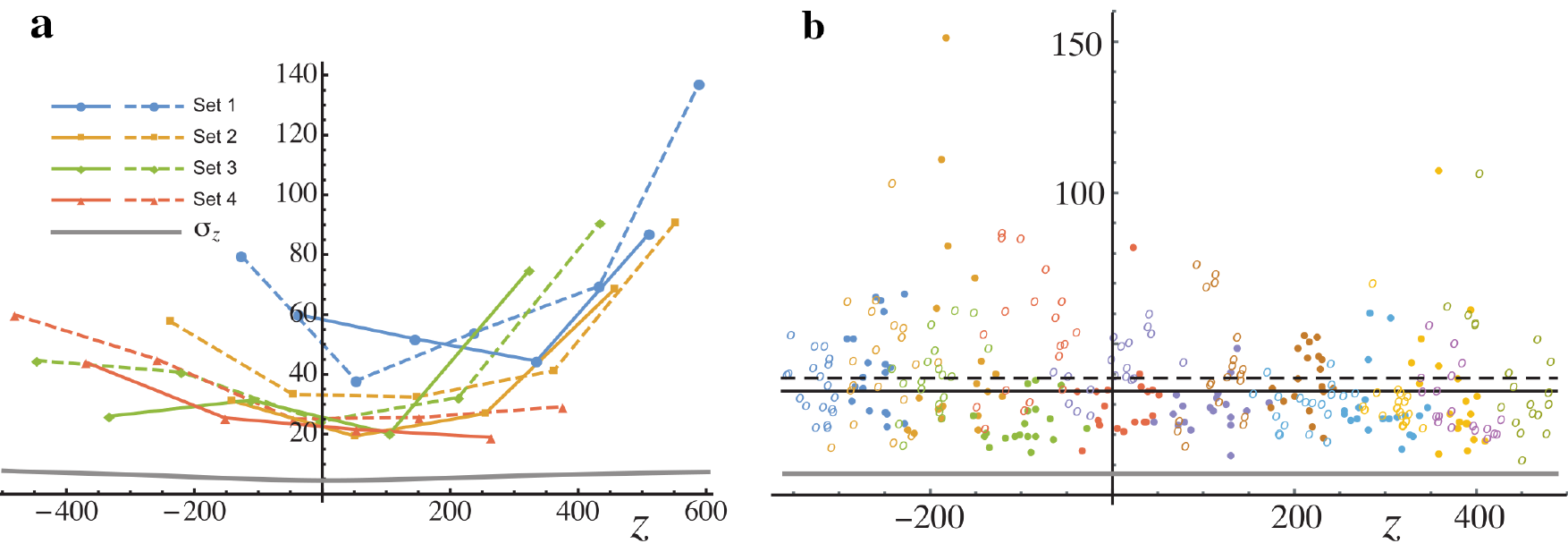}
\caption*{Supplementary Figure 10: Statistical analysis of the estimations of height for bead measurements. (a) For the four sets of bead measurements simulating molecules oriented in the $z$ direction, standard deviations of the $z$ estimate (dashed) and standard deviations of the $z$ increment per bead (solid), where the horizontal axis coordinate corresponds to the average of the $z$ estimates for the corresponding measurements for each set. The gray line corresponds to the CR lower bound calculated using the same model used for the retrieval, for 50000 photons and SBR = 3. (b) For the two sets of bead measurements simulating molecules oriented in several in-plane directions, standard deviations of the $z$ estimate (open dots) and standard deviations of the $z$ increment per bead (solid dots), where the horizontal axis coordinate corresponds to the average of the $z$ estimates for the corresponding measurements for each set. Each dot color represents the measurements for a given height and the different dots with the same color correspond to an image simulated a given molecule direction. The dashed and solid black lines represents the average spreads for the heights and height increments, respectively, while gray line corresponds to the maximum CR lower bound calculated using the same model used for the retrieval, for 50000 photons and SBR = 3.}
\label{spreadbeads}
\end{figure}

\subsection{Complete basis for fluorophores.}
The measurements with beads had the goal of showing the feasibility of height and in-plane orientation measurements. For the fluorophore measurements, on the other hand, the orientations are not known to be purely in-plane or out-of-plane, and in general there is wobbling. Therefore it is no longer possible to use a sub-basis; the complete basis of PSF elements is required.

For the first set of single molecule measurements (at different heights) the strategy we used was to combine the two sub-bases obtained from the bead reference PSFs. This information is not complete, so it had to be supplemented with theoretical calculations for conforming the basis, as explained in what follows. 
There were several challenges in combining the sub-bases obtained from the measurements using the S-waveplate and the linear polarizers. First, the two sets of reference measurements are not co-centered in the $xy$-plane, and in fact there is no guarantee that they are consistent in $z$ either because it is challenging experimentally to know the exact, absolute 3D position of a given bead. Second, the S-waveplate measurements provide distributions that are proportional to those appearing in the definitions of ${\cal I}_0^{(p)}$ and ${\cal I}_8^{(p)}$ in Eqs.~(\ref{PSFcomponents}a) and (\ref{PSFcomponents}i), namely $|G_z^{(p)}|^2$. However, the remaining elements of the PSF basis require cross terms between the transverse and axial components, and the rigorous experimental determination of these would require a reference dipole oriented, say, at $45^\circ$ from the $z$ axis, which is not easy to achieve experimentally. 

While there are other possible strategies for addressing this issue, the one used here was to combine the measurements with the theoretical model. Some parameters of the theoretical model were adjusted so that the theoretical predictions were as consistent as possible with the reference measurements. One parameter in particular was the orientation of the SEO, which was rotated by an angle of $-54.4^\circ$ with respect to the configuration corresponding to Eq.~(1) in the main text. Comparisons between theory and the experimental measurements allowed determining the relation between the $x$, $y$, and $z$ coordinates of the two reference sets. Centering with respect to $(x,y)$ was performed by multiplying by the appropriate linear phase factor in the Fourier domain. For each of the two sets, a quadratic fit in $z$ was performed, as described earlier, which allowed defining for each the nominal $z=0$ value for which the RHC and LHC PSFs are most aligned. The two sets of measurements were then renormalized to be as mutually consistent as possible when compared to the corresponding theoretical calculations. From these results, the magnitudes of $G_i^{(p)}$ for $i=x,y,z$ could be calculated by taking the square root of the corresponding component. The problem of the missing phases of $G_i^{(p)}$ required for the calculation of ${\cal I}_4^{(p)}$ and ${\cal I}_6^{(p)}$ was then resolved by using those from the adjusted theoretical models. The resulting PSF basis is shown in Supplementary Fig.~8(b). For comparison,  Supplementary Fig.~8(a) shows the theoretically-calculated PSF basis for the estimated SEO orientation. 

A different approach was used for the STORM measurements. Given the higher resolution of the camera used for those measurements as well as a higher value of $c=1.2\pi$, it was better not to use directly the PSFs measured for fluorescent beads because the blurring due to the bead's size was appreciable. Instead, the parameters of the theoretical model were adjusted so that, when blurred by the 3D dimension of the beads, they resembled those measured with beads. This theoretical model was then applied to retrieve the values of the parameters by maximizing their correlation with the measured PSFs. 
It should also be noted that, because the STORM measurements used a camera with smaller pixels, we used arrays of $29\times29$ pixels for the references instead of $21\times21$.

\begin{addendum}
 \item[Acknowledgements] This research has received funding from: National Science Foundation (NSF) (PHY-1507278); Excellence Initiative of Aix-Marseille Universit\'e (AMU) A*MIDEX, a French ``Investissements d'Avenir'' programme; European Union's Horizon 2020 research and innovation programme under the Marie Sklodowska-Curie grant agreement No 713750. Regional Council of Provence- Alpes-C\^ote d’Azur. A*MIDEX (No ANR- 11-IDEX-0001-02) from the Investissements d'Avenir project funded by the French Government, managed by the French National Research Agency (ANR). CONACYT Doctoral Fellowship. \\
 The authors are grateful to P. R\'efr\'egier who was instrumental in establishing this collaboration. They also thank A. J. Vella and M. Mavrakis, as well as A. M. Taddese, J. Puig and M. Rahman for help in the method development. Additionally, the authors thank the Center for Integrated Research Computing (CIRC) at the University of Rochester for providing computational resources.
\item[Author contributions] 
M.A.A. and S.B. conceived and initiated the project, inspired on a polarimetry method by T.G.B. who designed and provided the SEO. 
S.B., L.A.C. and V.C. designed and built the optical system. V.C. and L.A.C. prepared the samples and performed experiments. M.A.A wrote the algorithm, performed the theoretical developments and analyzed the data. L.A.C. and M.A.A. performed the Monte Carlo simulations. All authors wrote the paper and contributed to the scientific discussion and revisions.
\item[Author information] Correspondence and requests for materials should be addressed to M.A. Alonso (email: miguel.alonso@fresnel.fr) and S. Brasselet~(email: sophie.brasselet@fresnel.fr).
\end{addendum}


\newpage

\section*{References}
\begin{thebibliography}{10}
\expandafter\ifx\csname url\endcsname\relax
  \def\url#1{\texttt{#1}}\fi
\expandafter\ifx\csname urlprefix\endcsname\relax\def\urlprefix{URL }\fi
\providecommand{\bibinfo}[2]{#2}
\providecommand{\eprint}[2][]{\url{#2}}

\bibitem{Beausang2013}
\bibinfo{author}{Beausang, {\relax J. F}.}, \bibinfo{author}{Shroder, {\relax
  D. Y}.}, \bibinfo{author}{Nelson, {\relax P. C}.} \&
  \bibinfo{author}{Goldman, {\relax Y. E}.}
\newblock \bibinfo{title}{Tilting and wobble of myosin v by high-speed
  single-molecule polarized fluorescence microscopy}.
\newblock \emph{\bibinfo{journal}{Byopshys. J.}}
  \textbf{\bibinfo{volume}{104}}, \bibinfo{pages}{1263--1273}
  (\bibinfo{year}{2013}).

\bibitem{ValadesCruz2016}
\bibinfo{author}{Valad\'es~Cruz, {\relax C. A}.} \emph{et~al.}
\newblock \bibinfo{title}{{Quantitative nanoscale imaging of orientational
  order in biological filaments by polarized superresolution microscopy}}.
\newblock \emph{\bibinfo{journal}{Proc. Nat. Acad. Sci.}}
  \textbf{\bibinfo{volume}{113}}, \bibinfo{pages}{E820--E828}
  (\bibinfo{year}{2016}).

\bibitem{Mehta2017}
\bibinfo{author}{Mehta, {\relax S. B}.} \emph{et~al.}
\newblock \bibinfo{title}{{Dissection of molecular assembly dynamics by
  tracking orientation and position of single molecules in live cells}}.
\newblock \emph{\bibinfo{journal}{Proc. Nat. Acad. Sci.}}
  \textbf{\bibinfo{volume}{113}}, \bibinfo{pages}{E6352--E6361}
  (\bibinfo{year}{2016}).

\bibitem{Ding2020}
\bibinfo{author}{Ding, T.}, \bibinfo{author}{Wu, T.}, \bibinfo{author}{Mazidi,
  H.}, \bibinfo{author}{Zhang, O.} \& \bibinfo{author}{Lew, M.~D.}
\newblock \bibinfo{title}{Single-molecule orientation localization microscopy
  for resolving structural heterogeneities between amyloid fibrils}.
\newblock \emph{\bibinfo{journal}{Optica}} \textbf{\bibinfo{volume}{7}},
  \bibinfo{pages}{602--607} (\bibinfo{year}{2020}).

\bibitem{Backer2015}
\bibinfo{author}{Backer, {\relax A. S}.} \& \bibinfo{author}{Moerner, {\relax
  W. E}.}
\newblock \bibinfo{title}{{Determining the rotational mobility of a single
  molecule from a single image: a numerical study}}.
\newblock \emph{\bibinfo{journal}{Opt. Express}} \textbf{\bibinfo{volume}{23}},
  \bibinfo{pages}{4255--4276} (\bibinfo{year}{2015}).

\bibitem{Enderlein2006}
\bibinfo{author}{Enderlein, J.}, \bibinfo{author}{Toprak, E.} \&
  \bibinfo{author}{Selvin, {\relax P. R}.}
\newblock \bibinfo{title}{{Polarization effect on position accuracy of
  fluorophore localization}}.
\newblock \emph{\bibinfo{journal}{Opt. Express}} \textbf{\bibinfo{volume}{14}},
  \bibinfo{pages}{8111--8120} (\bibinfo{year}{2006}).

\bibitem{Backlund2012}
\bibinfo{author}{Backlund, {\relax M. P}.} \emph{et~al.}
\newblock \bibinfo{title}{{Simultaneous, accurate measurement of the 3D
  position and orientation of single molecules}}.
\newblock \emph{\bibinfo{journal}{Proc. Nat. Acad. Sci.}}
  \textbf{\bibinfo{volume}{109}}, \bibinfo{pages}{19087--92}
  (\bibinfo{year}{2012}).

\bibitem{Agrawal}
\bibinfo{author}{Agrawal, A.}, \bibinfo{author}{Quirin, S.},
  \bibinfo{author}{Grover, G.} \& \bibinfo{author}{Piestun, R.}
\newblock \bibinfo{title}{{Limits of 3D dipole localization and orientation
  estimation for single-molecule imaging: towards Green's tensor engineering}}.
\newblock \emph{\bibinfo{journal}{Opt. Express}} \textbf{\bibinfo{volume}{20}},
  \bibinfo{pages}{26667--26680} (\bibinfo{year}{2012}).

\bibitem{Backer2013}
\bibinfo{author}{Backer, {\relax A. S}.}, \bibinfo{author}{Backlund, {\relax M.
  P}.}, \bibinfo{author}{Lew, {\relax M. D}.} \& \bibinfo{author}{Moerner,
  {\relax W. E}.}
\newblock \bibinfo{title}{{Single-molecule orientation measurements with a
  quadrated pupil}}.
\newblock \emph{\bibinfo{journal}{Optics Letters}}
  \textbf{\bibinfo{volume}{38}}, \bibinfo{pages}{1521--1523}
  (\bibinfo{year}{2013}).

\bibitem{Backer2014}
\bibinfo{author}{Backer, {\relax A. S}.}, \bibinfo{author}{Backlund, {\relax M.
  P}.}, \bibinfo{author}{von Diezmann, {\relax A. R}.}, \bibinfo{author}{Sahl,
  {\relax S. J}.} \& \bibinfo{author}{Moerner, {\relax W. E}.}
\newblock \bibinfo{title}{{A bisected pupil for studying single-molecule
  orientational dynamics and its application to three-dimensional
  super-resolution microscopy}}.
\newblock \emph{\bibinfo{journal}{Appl. Phys. Lett.}}
  \textbf{\bibinfo{volume}{104}}, \bibinfo{pages}{193701}
  (\bibinfo{year}{2014}).

\bibitem{Mortensen2010}
\bibinfo{author}{Mortensen, {\relax K. I}.}, \bibinfo{author}{Churchman,
  {\relax L. S}.}, \bibinfo{author}{Spudich, {\relax J. A}.} \&
  \bibinfo{author}{Flyvbjerg, H.}
\newblock \bibinfo{title}{{Optimized localization analysis for single-molecule
  tracking and super-resolution microscopy}}.
\newblock \emph{\bibinfo{journal}{Nature Methods}}
  \textbf{\bibinfo{volume}{7}}, \bibinfo{pages}{377--381}
  (\bibinfo{year}{2010}).

\bibitem{Zhang2018}
\bibinfo{author}{Zhang, O.}, \bibinfo{author}{Lu, J.}, \bibinfo{author}{Ding,
  T.} \& \bibinfo{author}{Lew, {\relax M. D}.}
\newblock \bibinfo{title}{{Imaging the Three-Dimensional Orientation and
  Rotational Mobility of Fluorescent Emitters using the Tri-Spot Point Spread
  Function}}.
\newblock \emph{\bibinfo{journal}{Appl. Phys. Lett.}}
  \textbf{\bibinfo{volume}{113}}, \bibinfo{pages}{031103}
  (\bibinfo{year}{2018}).

\bibitem{Aguet2009}
\bibinfo{author}{Aguet, F.}, \bibinfo{author}{Geissb{\"u}hler, S.},
  \bibinfo{author}{Märki, I.}, \bibinfo{author}{Lasser, T.} \&
  \bibinfo{author}{Unser, M.}
\newblock \bibinfo{title}{{Super-resolution orientation estimation and
  localization of fluorescent dipoles using 3-D steerable filters}}.
\newblock \emph{\bibinfo{journal}{Optics Express}}
  \textbf{\bibinfo{volume}{17}}, \bibinfo{pages}{6829--6848}
  (\bibinfo{year}{2009}).

\bibitem{Corrie1999}
\bibinfo{author}{Corrie, J. E.~T.}, \bibinfo{author}{Brandmeier, B.~D.},
  \bibinfo{author}{Ferguson, R.~E.}, \bibinfo{author}{Trentham, D.~R.} \&
  \bibinfo{author}{al, e.}
\newblock \bibinfo{title}{Dynamic measurement of myosin light-chain-domain tilt
  and twist in muscle contraction}.
\newblock \emph{\bibinfo{journal}{Nature}} \textbf{\bibinfo{volume}{400}},
  \bibinfo{pages}{425--30} (\bibinfo{year}{1999}).

\bibitem{Sosa2001}
\bibinfo{author}{Sosa, H.}, \bibinfo{author}{Peterman, E.~J.},
  \bibinfo{author}{Moerner, W.~E.} \& \bibinfo{author}{Goldstein, L.~S.}
\newblock \bibinfo{title}{Adp-induced rocking of the kinesin motor domain
  revealed by single-molecule fluorescence polarization microscopy}.
\newblock \emph{\bibinfo{journal}{Nature structural biology}}
  \textbf{\bibinfo{volume}{8}}, \bibinfo{pages}{540--544}
  (\bibinfo{year}{2001}).

\bibitem{Peterman2001}
\bibinfo{author}{Peterman, E.~J.}, \bibinfo{author}{Sosa, H.},
  \bibinfo{author}{Goldstein, L.~S.} \& \bibinfo{author}{Moerner, W.}
\newblock \bibinfo{title}{Polarized fluorescence microscopy of individual and
  many kinesin motors bound to axonemal microtubules}.
\newblock \emph{\bibinfo{journal}{Biophysical Journal}}
  \textbf{\bibinfo{volume}{81}}, \bibinfo{pages}{2851 -- 2863}
  (\bibinfo{year}{2001}).

\bibitem{Backer2016}
\bibinfo{author}{Backer, {\relax A. S}.}, \bibinfo{author}{Lee, {\relax M. Y}.}
  \& \bibinfo{author}{Moerner, {\relax W. E}.}
\newblock \bibinfo{title}{{Enhanced DNA imaging using super-resolution
  microscopy and simultaneous single-molecule orientation measurements}}.
\newblock \emph{\bibinfo{journal}{Optica}} \textbf{\bibinfo{volume}{3}},
  \bibinfo{pages}{659--666} (\bibinfo{year}{2016}).

\bibitem{Roshita}
\bibinfo{author}{Ramkhalawon, R.}, \bibinfo{author}{Brown, {\relax T. G}.} \&
  \bibinfo{author}{Alonso, {\relax M. A}.}
\newblock \bibinfo{title}{{Imaging the polarization of a light field}}.
\newblock \emph{\bibinfo{journal}{Opt. Express}} \textbf{\bibinfo{volume}{21}},
  \bibinfo{pages}{4106--4115} (\bibinfo{year}{2013}).

\bibitem{Brandon}
\bibinfo{author}{Zimmerman, {\relax B. G}.}, \bibinfo{author}{Ramkhalawon, R.},
  \bibinfo{author}{Alonso, {\relax M. A}.} \& \bibinfo{author}{Brown, {\relax
  T. G}.}
\newblock \bibinfo{title}{{Pinhole array implementation of star test
  polarimetry}}.
\newblock \emph{\bibinfo{journal}{Proc. SPIE}} \textbf{\bibinfo{volume}{8949}},
  \bibinfo{pages}{894912} (\bibinfo{year}{2014}).

\bibitem{Brandon2}
\bibinfo{author}{Zimmerman, {\relax B. G}.} \& \bibinfo{author}{Brown, {\relax
  T. G}.}
\newblock \bibinfo{title}{{Star test image-sampling polarimeter}}.
\newblock \emph{\bibinfo{journal}{Opt. Express}} \textbf{\bibinfo{volume}{24}},
  \bibinfo{pages}{23154--23161} (\bibinfo{year}{2016}).

\bibitem{BrandonAerosols}
\bibinfo{author}{Zimmerman, {\relax B. G}.}, \bibinfo{author}{Adamson, P.} \&
  \bibinfo{author}{Brown, {\relax T. G}.}
\newblock \bibinfo{title}{{Exploring new polarimetric techniques using
  unconventionally polarized sources}}.
\newblock \emph{\bibinfo{journal}{Proc. SPIE}} \textbf{\bibinfo{volume}{8515}},
  \bibinfo{pages}{85150Q} (\bibinfo{year}{2012}).

\bibitem{Sid}
\bibinfo{author}{Sivankutty, S.} \emph{et~al.}
\newblock \bibinfo{title}{{Single-shot polarimetry imaging of multicore
  fiber}}.
\newblock \emph{\bibinfo{journal}{Opt. Lett.}} \textbf{\bibinfo{volume}{41}},
  \bibinfo{pages}{2105--2108} (\bibinfo{year}{2016}).

\bibitem{Alexis}
\bibinfo{author}{Spilman, {\relax A. K}.} \& \bibinfo{author}{Brown, {\relax T.
  G}.}
\newblock \bibinfo{title}{{Stress birefringent, space-variant wave plates for
  vortex illumination}}.
\newblock \emph{\bibinfo{journal}{Appl. Opt.}} \textbf{\bibinfo{volume}{26}},
  \bibinfo{pages}{61--66} (\bibinfo{year}{2007}).

\bibitem{Anthony}
\bibinfo{author}{Vella, {\relax A. J}.} \& \bibinfo{author}{Alonso, {\relax M.
  A}.}
\newblock \bibinfo{title}{{Optimal birefringence distributions for imaging
  polarimetry}}.
\newblock \emph{\bibinfo{journal}{Opt. Express}} \textbf{\bibinfo{volume}{27}},
  \bibinfo{pages}{36799--36814} (\bibinfo{year}{2019}).

\bibitem{Bohmer:03}
\bibinfo{author}{B\"{o}hmer, M.} \& \bibinfo{author}{Enderlein, J.}
\newblock \bibinfo{title}{Orientation imaging of single molecules by wide-field
  epifluorescence microscopy}.
\newblock \emph{\bibinfo{journal}{J. Opt. Soc. Am. B}}
  \textbf{\bibinfo{volume}{20}}, \bibinfo{pages}{554--559}
  (\bibinfo{year}{2003}).

\bibitem{HieuThao:20}
\bibinfo{author}{{\relax Hieu Thao}, N.}, \bibinfo{author}{Soloviev, O.} \&
  \bibinfo{author}{Verhaegen, M.}
\newblock \bibinfo{title}{Phase retrieval based on the vectorial model of point
  spread function}.
\newblock \emph{\bibinfo{journal}{J. Opt. Soc. Am. A}}
  \textbf{\bibinfo{volume}{37}}, \bibinfo{pages}{16--26}
  (\bibinfo{year}{2020}).

\bibitem{Chandler1:19}
\bibinfo{author}{Chandler, T.}, \bibinfo{author}{Shroff, H.},
  \bibinfo{author}{Oldenbourg, R.} \& \bibinfo{author}{{\relax La Rivi\`{e}re},
  P.}
\newblock \bibinfo{title}{Spatio-angular fluorescence microscopy i. basic
  theory}.
\newblock \emph{\bibinfo{journal}{J. Opt. Soc. Am. A}}
  \textbf{\bibinfo{volume}{36}}, \bibinfo{pages}{1334--1345}
  (\bibinfo{year}{2019}).

\bibitem{Chandler2:19}
\bibinfo{author}{Chandler, T.}, \bibinfo{author}{Shroff, H.},
  \bibinfo{author}{Oldenbourg, R.} \& \bibinfo{author}{{\relax La Rivi\`{e}re},
  P.}
\newblock \bibinfo{title}{Spatio-angular fluorescence microscopy ii. paraxial
  4f imaging}.
\newblock \emph{\bibinfo{journal}{J. Opt. Soc. Am. A}}
  \textbf{\bibinfo{volume}{36}}, \bibinfo{pages}{1346--1360}
  (\bibinfo{year}{2019}).

\bibitem{Brosseau}
\bibinfo{author}{Brosseau, C.}
\newblock \emph{\bibinfo{title}{{Fundamentals of Polarized Light}}}
  (\bibinfo{publisher}{John Wiley \& Sons, Inc.}, \bibinfo{year}{1998}).

\bibitem{Sampson}
\bibinfo{author}{Samson, {\relax J. C}.}
\newblock \bibinfo{title}{{Descriptions of the polarization states of vector
  processes: applications to ULF magnetic fields}}.
\newblock \emph{\bibinfo{journal}{Geophys. J. R. Astron. Soc.}}
  \textbf{\bibinfo{volume}{34}}, \bibinfo{pages}{403--419}
  (\bibinfo{year}{1973}).

\bibitem{Barakat}
\bibinfo{author}{Barakat, R.}
\newblock \bibinfo{title}{{Degree of polarization and the principal idempotents
  of the coherency matrix}}.
\newblock \emph{\bibinfo{journal}{Opt. Commun.}} \textbf{\bibinfo{volume}{23}},
  \bibinfo{pages}{147--150} (\bibinfo{year}{1977}).

\bibitem{Tero}
\bibinfo{author}{Set{\"a}l{\"a}, T.}, \bibinfo{author}{Shevchenko, A.},
  \bibinfo{author}{Kaivola, M.} \& \bibinfo{author}{Friberg, {\relax A. T}.}
\newblock \bibinfo{title}{{Degree of polarization for optical near fields}}.
\newblock \emph{\bibinfo{journal}{Phys. Rev. E}} \textbf{\bibinfo{volume}{66}},
  \bibinfo{pages}{016615} (\bibinfo{year}{2002}).

\bibitem{2methods}
\bibinfo{author}{Petruccelli, {\relax J. C}.}, \bibinfo{author}{Moore, {\relax
  N. J}.} \& \bibinfo{author}{Alonso, {\relax M. A}.}
\newblock \bibinfo{title}{{Two methods for modeling the propagation of the
  coherence and polarization properties of nonparaxial fields}}.
\newblock \emph{\bibinfo{journal}{Opt. Commun.}}
  \textbf{\bibinfo{volume}{283}}, \bibinfo{pages}{4457--4466}
  (\bibinfo{year}{2010}).

\bibitem{Alonso:2020geometric}
\bibinfo{author}{Alonso, {\relax M. A}.}
\newblock \bibinfo{title}{Geometric descriptions for the polarization for
  nonparaxial optical fields: a tutorial}.
\newblock \bibinfo{journal}{Preprint at http://arxiv.org/abs/2008.02720} 
  (\bibinfo{year}{2020}).

\bibitem{Zhang2019}
\bibinfo{author}{Zhang, O.} \& \bibinfo{author}{Lew, M.~D.}
\newblock \bibinfo{title}{Fundamental limits on measuring the rotational
  constraint of single molecules using fluorescence microscopy}.
\newblock \emph{\bibinfo{journal}{Phys. Rev. Lett.}}
  \textbf{\bibinfo{volume}{122}}, \bibinfo{pages}{198301}
  (\bibinfo{year}{2019}).

\bibitem{Ober2004}
\bibinfo{author}{Ober, {\relax R. J}.}, \bibinfo{author}{Ram, S.} \&
  \bibinfo{author}{{\relax E. S}, W.}
\newblock \bibinfo{title}{Localization accuracy in single-molecule microscopy}.
\newblock \emph{\bibinfo{journal}{Biophysical Journal}}
  \textbf{\bibinfo{volume}{86}}, \bibinfo{pages}{1185--1200}
  (\bibinfo{year}{2004}).

\bibitem{Vella2020}
\bibinfo{author}{Vella, A.} \& \bibinfo{author}{Alonso, {\relax M. A}.}
\newblock \bibinfo{title}{Maximum likelihood estimation in the context of an
  optical measurement}.
\newblock \emph{\bibinfo{journal}{Progress in Optics}}
  \textbf{\bibinfo{volume}{65}}, \bibinfo{pages}{231--311}
  (\bibinfo{year}{2020}).

\bibitem{Dempsey2011}
\bibinfo{author}{Dempsey, {\relax G. T}.}, \bibinfo{author}{Vaughan, {\relax J.
  C}.}, \bibinfo{author}{Chen, {\relax K. H}.}, \bibinfo{author}{Bates, M.} \&
  \bibinfo{author}{Zhuang, X.}
\newblock \bibinfo{title}{{Evaluation of fluorophores for optimal performance
  in localization-based super-resolution imaging}}.
\newblock \emph{\bibinfo{journal}{Nature methods}}
  \textbf{\bibinfo{volume}{8}}, \bibinfo{pages}{1027--36}
  (\bibinfo{year}{2011}).

\bibitem{Mark}
\bibinfo{author}{Dennis, {\relax M. R}.}
\newblock \bibinfo{title}{{Geometric interpretation of the three-dimensional
  coherence matrix for nonparaxial polarization}}.
\newblock \emph{\bibinfo{journal}{J. Opt. A: Pure Appl. Opt.}}
  \textbf{\bibinfo{volume}{6}}, \bibinfo{pages}{228--231}
  (\bibinfo{year}{2004}).

\bibitem{Holden2011}
\bibinfo{author}{Holden, {\relax S. J}.}, \bibinfo{author}{Uphoff, S.} \&
  \bibinfo{author}{Kapanidis, {\relax A. N}.}
\newblock \bibinfo{title}{{DAOSTORM: an algorithm for high- density
  super-resolution microscopy}}.
\newblock \emph{\bibinfo{journal}{Nature Methods}}
  \textbf{\bibinfo{volume}{8}}, \bibinfo{pages}{279--280}
  (\bibinfo{year}{2011}).

\bibitem{Huang2011}
\bibinfo{author}{Huang, F.}, \bibinfo{author}{Schwartz, {\relax S. L}.},
  \bibinfo{author}{Byars, {\relax J. M}.} \& \bibinfo{author}{Lidke, {\relax K.
  A}.}
\newblock \bibinfo{title}{{Simultaneous multiple-emitter fitting for single
  molecule super-resolution imaging}}.
\newblock \emph{\bibinfo{journal}{Biomed. Opt. Express}}
  \textbf{\bibinfo{volume}{2}}, \bibinfo{pages}{1377--1393}
  (\bibinfo{year}{2011}).

\bibitem{Zhu2012}
\bibinfo{author}{Zhu, L.}, \bibinfo{author}{Zhang, W.},
  \bibinfo{author}{Elnatan, D.} \& \bibinfo{author}{Huang, B.}
\newblock \bibinfo{title}{{Faster STORM using compressed sensing}}.
\newblock \emph{\bibinfo{journal}{Nature Methods}}
  \textbf{\bibinfo{volume}{9}}, \bibinfo{pages}{721--723}
  (\bibinfo{year}{2012}).

\bibitem{Mailfert2018}
\bibinfo{author}{Mailfert, S.} \emph{et~al.}
\newblock \bibinfo{title}{{Theoretical High-Density Nanoscopy Study Leads to
  the Design of UNLOC, a Parameter-free Algorithm}}.
\newblock \emph{\bibinfo{journal}{Biophysical Journal}}
  \textbf{\bibinfo{volume}{115}}, \bibinfo{pages}{565--576}
  (\bibinfo{year}{2018}).

\bibitem{Barsic2015}
\bibinfo{author}{Barsic, A.}, \bibinfo{author}{Grover, G.} \&
  \bibinfo{author}{R., P.}
\newblock \bibinfo{title}{{Three-dimensional super-resolution and localization
  of dense clusters of single molecules}}.
\newblock \emph{\bibinfo{journal}{Scientific Reports}}
  \textbf{\bibinfo{volume}{4}} (\bibinfo{year}{2015}).

\bibitem{Mazidi2019}
\bibinfo{author}{{Mazidi}, H.}, \bibinfo{author}{{King}, {\relax E. S}.},
  \bibinfo{author}{{Zhang}, O.}, \bibinfo{author}{{Nehorai}, A.} \&
  \bibinfo{author}{{Lew}, M.~D.}
\newblock \bibinfo{title}{Dense super-resolution imaging of molecular
  orientation via joint sparse basis deconvolution and spatial pooling}.
\newblock In \emph{\bibinfo{booktitle}{2019 IEEE 16th International Symposium
  on Biomedical Imaging (ISBI 2019)}}, \bibinfo{pages}{325--329}
  (\bibinfo{year}{2019}).

\bibitem{Lindfors}
\bibinfo{author}{Lindfors, K.} \emph{et~al.}
\newblock \bibinfo{title}{{Local polarization of tightly focused unpolarized
  light}}.
\newblock \emph{\bibinfo{journal}{Nature Photonics}}
  \textbf{\bibinfo{volume}{1}}, \bibinfo{pages}{147--150}
  (\bibinfo{year}{2007}).

\bibitem{Lukas}
\bibinfo{author}{Deutsch, B.}, \bibinfo{author}{Hillenbrand, R.} \&
  \bibinfo{author}{Novotny, L.}
\newblock \bibinfo{title}{{Visualizing the Optical Interaction Tensor of a Gold
  Nanoparticle Pair}}.
\newblock \emph{\bibinfo{journal}{Nano Lett.}} \textbf{\bibinfo{volume}{10}},
  \bibinfo{pages}{652--656} (\bibinfo{year}{2010}).

\bibitem{Spilmanthesis}
\bibinfo{author}{Spilman, A.}
\bibinfo{title}{\emph{Stress-engineered optical elements}, Doctoral dissertation, University of Rochester, 2007.}

\bibitem{Beckleythesis}
\bibinfo{author}{Beckley, A.M.}
\bibinfo{title}{\emph{Polarimetry and beam apodization using stress-engineered optical elements}, Doctoral dissertation, University of Rochester, 2012.}


\end{thebibliography}

\end{document}